\documentclass[]{aa} 
\usepackage{amsmath}
\usepackage{txfonts}

\usepackage[utf8]{inputenc}

\usepackage{boxedminipage}

\usepackage{listings}

\usepackage{url}

\usepackage{tikz}
\usetikzlibrary{snakes}

\usepackage{natbib}
\bibpunct{(}{)}{;}{a}{}{,} 

\usepackage{slashbox} 

\begin{document}
   \title{Statistical techniques for the detection and analysis of
     solar explosive events}
   \titlerunning{Statistical analysis of explosive events}
   \author{
    L.M. Sarro
     \inst{1}
     \and
     A. Berihuete
     \inst{2}
   }
   \institute{
     Dpt. de Inteligencia Artificial , UNED, Juan del Rosal, 16, 28040
     Madrid, Spain\\ 
     \and
     Dpt. Statistics and Operations Research, University of
     C\'adiz, Campus Universitario R\'io San Pedro s/n.  11510 Puerto
     Real, C\'adiz, Spain\\ 
   }
       \date{}
       
 
  \abstract
      {Solar explosive events are commonly explained as small scale
        magnetic reconnection events, although unambiguous
        confirmation of this scenario remains elusive due to the lack
        of spatial resolution and the statistical analysis of large
        enough samples of this type of events.}
   {In this work, we propose a sound statistical treatment of data
     cubes consisting of a temporal sequence of long slit spectra of
     the solar atmosphere. The analysis comprises all the stages from
     the explosive event detection to its characterization and the
     subsequent sample study.}
  {We have designed two complementary approaches based on the
    combination of standard statistical techniques (Robust Principal
    Component Analysis in one approach and wavelet decomposition and
    Independent Component Analysis in the second) in order to obtain
    least biased samples. These techniques are implemented in the
    spirit of letting the data speak for themselves. The analysis is
    carried out for two spectral lines: the C~{\sc iv} line at 1548.2
    \AA~ and the Ne~{\sc viii} line at 770.4 \AA.}
   {We find significant differences between the characteristics of the
     line profiles emitted in the proximities of two active regions,
     and in the quiet Sun, most visible in the relative importance of a
     separate population of red shifted profiles. We also find a higher
     frequency of explosive events near the active regions, and in the
     C~{\sc iv} line. The distribution of the explosive events
     characteristics is interpreted in the light of recent numerical
     simulations. Finally, we point out several regions of the
     parameter space where the reconnection model has to be refined in
     order to explain the observations.}
   {}


   \maketitle
%

\section{Introduction}

Explosive events are localised energy release episodes detected mainly
as broad emission lines in solar transition region lines. They were
first discovered and classified by \cite{1983ApJ...272..329B} using
the {\sl High Resolution Telescope and Spectrograph}, HRTS. Their
properties, were summarized by \cite{1989SoPh..123...41D} and
\cite{1994AdSpR..14...13D}. 

\citep{1995SoPh..162..189W} has provided a wealth of detailed
observations in a wavelength range overlapping that of the HRTS, no
update of the statistical picture has been carried out based on the
new data. SUMER observations have been used in combination with other
SOHO and Earth-based instruments to explore the relationship of
explosive events with the magnetic field evolution
\citep{1998ApJ...497L.109C, Teriaca:2004a, 2003A&A...403..731M}; to
provide a coherent picture of explosive events in relation with other
transient events such as blinkers and/or surges
\citep{2009ApJ...701..253M, 2005A&A...432..307B, 2003A&A...403..731M,
  2003A&A...403..287P, 2000ApJ...528L.119C, 1998ApJ...504L.123C}; and
to explore specific aspects of the explosive event phenomenon, like
the timing and variations in lines of different formation temperatures
in \citet{2005A&A...431..339M}, and the comparison of the signatures
of explosive events in various lines of ions with similar formation
temperatures in \citet{2005A&A...439.1183D}.

The theoretical work developed in the past decade in order to explain
the observed non-gaussianity of the explosive events line profiles has
converged in the framework of magnetic reconnection
\citep{1997Natur.386..811I}. Recent examples of numerical simulations
of explosive events in this scenario can be found in
\citet{2010A&A...510A.111D}, \citet{2009ApJ...702....1H},
\citet{2009A&A...495..953L}, and \citet{2006SoPh..238..313C}.

In this paper, we propose an automatized procedure for the detection
and analysis of explosive events. We aim at studying their properties
in sufficiently large samples, and compare them with predictions from
the models, in the hope that, by pointing at the discrepancies between
observations and the numerical simulations, we can help refine the
models of magnetic reconnection. The outline of this paper is as
follows: in Sect. \ref{obs} we describe the observations used to
test the validity of the techniques proposed in Sect.
\ref{sec:methodology} for the detection and analysis of the explosive
events line profiles; in Sect. \ref{sec:results} we describe the results
of applying these techniques to the SUMER data, and in Sect.
\ref{discussion}, we discuss the general properties of the explosive
events samples thus obtained, and the match between the observed
properties and the simulations. 

\section{Observations}
\label{obs}

The observations used for the scientific validation of the techniques
outlined in the next section correspond to the study
VIA~JOP38\_MAY2000\_4L as indexed in the SOHO archive accessible via
web either at ESAC\footnote{\url{http://soho.esac.esa.int/}} or the
GSFC\footnote{\url{http://sohowww.nascom.nasa.gov/data/archive/index_gsfc.html}}.

They consist of two series of 60 seconds exposures taken with the
$1''\times 300''$ slit starting on May 18th, 2000 at 09:45:15 (first
series) and May 19th at 03:55:30 (second series) and ending at
14:47:16 (first series) and 08:57:31 (second series). All along the
observational sequence the slit was held fixed (the so-called Temporal
Serial sequence) centred in the equator and displaced $290''$ eastward
from the central meridian. The spatial resolution is $0.96''$ and the
spectral resolution is 0.04~\AA.

Since the compensation for solar rotation is disabled in the Temporal
Serial observational sequences, two consecutive exposures do not
exactly correspond to the same region of the Sun. At the centre of the
slit (which is at a declination of -290$''$) the motion of a plasma
element in the solar surface due to rotation is roughly $10.15''$ per
hour, or 0.17 arcsecs per minute. Therefore, the overlapping area
covered at the beginning of two consecutive exposures has a width of
$1-0.17=0.83$ arcsecs. 

The observations were reduced using the standard pipeline procedure
sum\_read\_corr\_fts.pro which amongst others, applies the flatfield,
deadtime, pixel linearity and distortion corrections and calibrates
the spectral images in wavelength and flux. The flatfield correction
resolves a small shift of the image caused by the channel plate. We
use the flatfield correction matrix closest to the time of the
observations~\footnote{\url{http://sohowww.nascom.nasa.gov/sdb/soho/sumer/calibration/flight/ff/}}
(ff\_a\_990311.fits). The detected intensity is given in W m$^2$
sr$^{-1}$ \AA$^{-1}$ as obtained with the calibration routine
radiometry.pro and the radiometric calibration\footnote{Files with the
  calibrations curves are obtained from
  ~\url{http://sohowww.nascom.nasa.gov/solarsoft/soho/sumer/idl/contrib/wilhelm/rad/}}.

The result of this data reduction are two three-dimensional data cubes
defined by three axes: the position along the slit, the dispersion
direction and time. We will often refer to one single spectral line
(at a given position along the slit and time) as a raster. Since the
observations were obtained with the slit position on the disk fixed,
there is no possible confusion with the so--called raster scans used
to cover a large area of the Sun by moving the slit in a definite
direction.

\begin{figure}[htp]
   \centering
   \includegraphics[scale=.20]{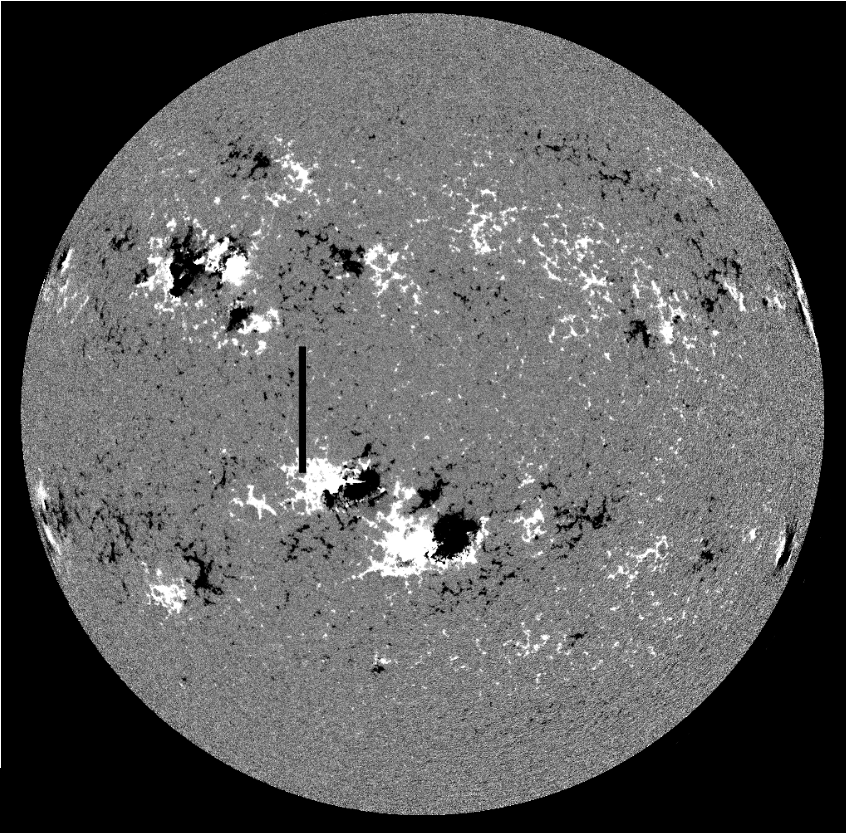}
   \caption{MDI magnetogram of the Sun taken on May 18th at 11:12
     UT. The vertical black line marks the approximate position of the
     SUMER slit in the first time series of spectral images.}
   \label{mdi18}
\end{figure}

\begin{figure}[htp]
   \centering
   \includegraphics[scale=.20]{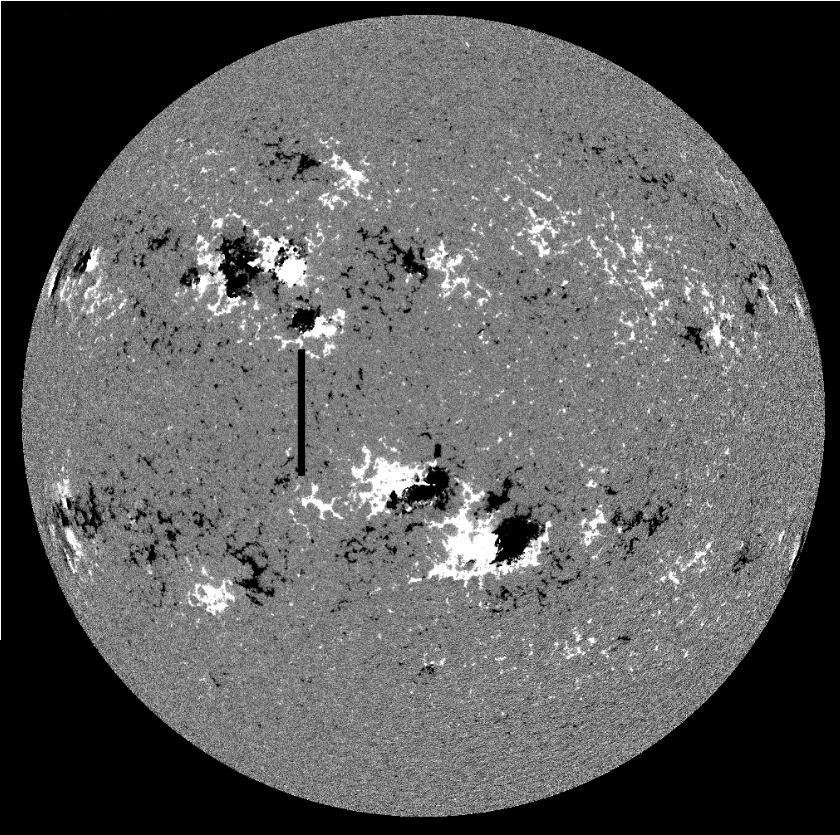}
   \caption{MDI magnetogram of the Sun taken on May 19th at 04:48
     UT. The vertical black line marks the approximate position of the
     SUMER slit in the first time series of spectral images.}
   \label{mdi19}
\end{figure}

Figures \ref{mdi18} and \ref{mdi19} show two MDI magnetograms obtained
on May 18th at 11:12 UT and on May 19th at 04:48 UT. They thus
correspond to intermediate times in the two series of
observations. Although the header information indicates that they
correspond to quiet Sun observations, it is evident that, at least the
first observational sequence is partly affected by active region NOAA
08998. We also find hints that the second time series is affected by
the external regions of NOAA 09004. In Sect. \ref{sec:results} we
will analyse separately these quiet Sun and active areas.

\section{Methodology} 
\label{sec:methodology}
The detection of explosive events recorded in a spectral image has
been tackled in the past using differents methods. A precise
description of these methos can be found in~\cite{Perez:1999a},
\cite{Ning:2004a}, \cite{Teriaca:2004a}, and \cite{Doyle:2006a}. All
these procedures are applied only to a small number of rasters (of the
order of a few hundreds), and in most cases the techniques require
visual inspection to decide whether a line profile is an explosive
event or not. Furthermore, they often consist in hard thresholding the
widths or integrated emission of a given spectral line (i.e., they
define an explosive event as a line profile with a line width
$n\sigma$ above the mean, with $n$ usually 3, and/or equivalently for
the line integrated emission). This work presents another perspective
based on the idea to allow the data \emph{speak for themselves}: we
implement automatic robust techiques to find outliers in a point
distribution where every line profile is characterized by a set of
properties parametrizing its non-gaussianity.


Throughout this section we will use the following notation: matrices
are denoted by capital letters, often with two subindices that
indicate the dimensions of the matrix. For instance, \( X_{n,p} \)
stands for a matrix with \( n \) rows and \( p \) columns, where \( n
\) indexes line profiles and \( p \) is the number of variables used
to characterise the line profile. These variables are further
specified in Sect. \ref{sec:results}, but the methodology described
below is independent of the choice of variables. A vector is
represented by boldface lowercase letter, and it is always assumed to
be a column vector. A vector \( \mathbf{u} \) with \( p \) components
is written as \( \mathbf{u}=(u_1, \dots , u_p)^T \).

\subsection{The PCOut algorithm} 
\label{sub:the_algorithm_pcout}

  Let $\mathbf{X}$ be a $n\times p$ matrix with each row containing
  one observation defined by $p$ variables. In our case, each row
  represents a line profile with $p$ flux values. The PCOut algorithm
  of \citet{Filzmoser:2008a} combines two complementary measures of
  outlyingness (location and scatter outlyingness) to provide a final
  score that allows the ranking of the observations (the line profiles
  in our case) according to their deviation from
  \emph{typicality}. Observe that \emph{typicality} refers to quiet
  sun profiles, and outilers to quiet sun profiles affected by
  explosive events. That is, letting \( F_1, \) the distribution of
  the non-outliers and \( F_2 \) the distribution of the outliers, the
  distribution of the dataset as a whole is usually considered to be
  given by \( F = (1 - \epsilon) F_1 + \epsilon F_2 \), for \( 0 \leq
  \epsilon <0.5 \). In our problem, \( F_1 \) and \( F_2 \) do not
  differ so much (disturbances in the profile), and we refer to
  \emph{location outlier} if the mean of distribution \( F_2 \) is
  different from \( F_1 \), and \emph{scatter outlier} if the variance
  of \( F_2 \) is different from \( F_1 \).

In the initial step, the algorithm normalises the data using the
L1-median as an estimate of the mean and the median absolute deviation
(MAD) as an estimate of the standard deviation. The L1-median is a
highly robust and orthogonally equivariant\footnote{Linear
  translations of the data are paralleled by a similar translation of
  the estimator} location estimator, and it is defined as the point \(
\theta \) which minimizes the sum of distances to all observations,
i.e.

\[
\boldsymbol{\mu}=\text{L1MED } ( \mathbf{x}_1, \dots , \mathbf{x}_n ) =
\underset{\boldsymbol{\theta}}{\text{argmin }}\, \sum_{i=1}^n \Vert
\mathbf{x}_i -\mathbf{\boldsymbol{\theta}} \Vert,
\]

where \( \Vert \cdot \Vert \) stands for the euclidean norm. The MAD
is defined for a sample \( \{ x_1, \dots , x_n \} \subset
\varmathbb{R} \) as

\[
\text{MAD } ( x_1, \dots , x_n ) = 1.4826 \cdot
\underset{j}{\text{median }} |x_j -\underset{i}{\text{median }} x_i |
\]

Since the MAD is used here as an estimator of the standard
  deviation $\sigma$, we need to introduce the constant scale factor
  1.4826 for consistency. The scale factor ensures that
\[
E[MAD(X_1, \dots, X_n )] = \sigma,
\]
for a random variable \( X\) distributed normally as \( N( \mu ,
\sigma^2 ) \) and large \( n \). 

We then compute the sample covariance matrix of the normalised
data, $\boldsymbol{C}$, and obtain the principal components of the
dataset by selecting the first \( q \quad (q<p)\) eigenvectors
that contribute at least a preselected percentage (in our case the
99\%) of the total variance. From the \( q \times q \) matrix \(
\mathbf{Q} \) of eigenvectors of the sample covariance matrix, we obtain
the principal components as \[ \mathbf{Z=X \cdot Q}\]

These principal components are rescaled by the median and the MAD according to,

\[
	z^{*}_{ij}= \frac{z_{ij}-\text{med}(z_{1j}, \dots , z_{nj})}{\text{MAD } (z_{1j}, \dots , z_{nj})}
\]

\subsubsection{Detecting location outliers} 
\label{ssub:detecting_location_outliers}

The location of the outliers begins by calculating the absolute value
of a robust kurtosis measure for each component:

\begin{equation}\label{eq:robust_kurtosis}
w^{*}_j= \left| 
\frac{1}{n} 
\left( 
\sum_{i=1}^{n} 
\frac
{(z^{*}_{ij}- \text{med}(z^{*}_{1j}, \dots , z^{*}_{nj}))^4}
{ \text{MAD }( z^{*}_{1j}, \dots , z^{*}_{nj})^4} 
\right)
-3 
\right|,
\end{equation}
where $z^{*}_{ij}$ is $j$-th coefficient of the $i$-th line profile in the
new basis of principal components, and \( j= 1, \dots , q \). In
practice we use relative weights \( w^{*}_j / \sum_i w^{*}_i \) to produce a
standard scale \( 0 \leq w^{*}_j \leq 1 \).

Equation~(\ref{eq:robust_kurtosis}) assigns higher weights to the
components where outliers clearly stand out. If no outliers are
present in a given component $j$ then we expect the kurtosis to be
close to 0, and \( w^{*}_j \approx 0 \). Note that principal components
are sorted in decreasing order of explained variance, but also that
outliers increase the variance along their respective
directions. Therefore: (i) outliers projected over principal
components space will be more visible than in the original space, and
(ii) we expect outliers to stand out clearly in one principal
component rather than being slightly apparent in all of them.

The robust Mahalanobis distances are computed taking into account the
weights \( w^{*}_j \) according to

\begin{equation}
RD_i=\sqrt{\sum_{j} (z^{*}_{ij} \cdot w^{*}_j)^2},
\end{equation}

The first phase of the algorithm continues with the transformation of
the $RD_i$ distances according to

\begin{equation}\label{eq:transf_distances}
d_i = RD_i \cdot \frac{\sqrt{\chi^2_{q,0.5}}}{\text{median}\{RD_i\}},
\end{equation}

where \( \chi^2_{q,0.5} \) is the \( 0.5 \) quantile of \( \chi^2_q
\). The transformation of the robust distances is needed because any
resemblance of the $d_i$ distribution with a $\chi^2_q$ distribution
is lost with the kurtosis weighting scheme. The distances \( d_i \)
are used to assign weights to each observation by means of the
biweight function:
\begin{equation}\label{eq:biweight_function}
w_{1i}=
\begin{cases}
	0, & d_i \geq c \\
	\left( 1-\left( \frac{d_i-M}{c-M}\right)^2\right)^2, & M <d_i<c \\
	1 & d_i \leq M
\end{cases}
\end{equation}
where \( i = 1, \dots , n \), \( M \) is the \( 0.97 \) quantile of
the distances \( \{ d_1, \dots ,d_n\} \) and \( c \) verifies:

\[
	c=\text{median} \{ d_1, \dots ,d_n\} + 10 \cdot \text{MAD } (
        d_1, \dots d_n ) .
\]

These weights are used as a measure of location outlyingness.

Note that the values of $M$ and $c$ are not the ones recommended in
\cite{Filzmoser:2008a}, but the values that we found to better
represent the distribution of SUMER line profiles described above.


\subsubsection{Detecting scatter outliers} 
\label{ssub:detecting_scatter_outliers}

The second phase of the algorithm is aimed at detecting the so-called
scatter outliers, i.e., outliers that do not stand out clearly in one
principal component, but that are slightly visible in many of
them. Using the same semi-robust principal components decomposition
obtained in the first stage of the algorithm, the euclidean norm for
data in principal component space is calculated. These distances are
equivalent to the Mahalanobis distance in the original data space, and
have not been transformed using the kurtosis weighting scheme defined
in Eq. (\ref{eq:robust_kurtosis}). Hence, the transformation in
Eq. (\ref{eq:transf_distances}) yields a distribution much closer to
\( \chi^2_q \). The biweight functions in
Eq.~(\ref{eq:biweight_function}) are set-up according to the results
in \citet{Filzmoser:2008a}, i.e. \( M^2 = \chi^2_{q,0.25} \) and \(
c^2 = \chi^2_{q,0.99} \), except that we need to used quantiles other
than 0.25 and 0.99. The values of these quantiles have to be much
closer to one in order to better represent a data distribution like
ours, where the number of outliers is much smaller that the group of
cases defining normality. The weights (measuring scatter outlyingness)
calculated in this step are called \( w_{2i}, \quad i=1, \dots , n\).


\subsubsection{Computation of final weights} 
\label{ssub:computation_of_final_weights}
Finally, the results of the two phases of the algorithm are combined,
and final weights \( w_i , \, i=1, \dots , n \) are calculated in
accordance with:
\[
w_i = \frac{(w_{1i}+s)(w_{2i}+s)}{(1+s)^2} ,
\]
where typically the scaling constant \( s=0.25 \). Outliers are
  then clasified as points having weights \( w_i < 0.25 \).



\subsection{Wavelet enhanced ICA quiet Sun removal} 
\label{sec:wavelet_enhanced_ica_artifact_removal}

Once the explosive events are identified by the PCOut algorithm, we
want to isolate the explosive event contribution to the line profile
from the surrounding quiet Sun emission.

In principle, it can be expected that a line profile revealing an
explosive event includes contributions from both the explosion site
and the surroundings, in a proportion which is related to the
respective volume emission measures and the filling factor. In this
section we describe a technique aimed at separating the two
contributions.

The Independent Component Analysis (ICA) is a statistical technique
for separating a multivariate signal into additive subcomponents,
assuming the mutual statistical independence of the non-Gaussian
source signals. It can be seen as a special case of the \emph{blind
  source separation problem}.

In contrast to Principal Components Analysis, ICA is based on the
three following assumptions: (i) the sources are statistically
independent; (ii) the probability densities of the sources are
non-Gaussian (at most one of them is allowed to be Gaussian); (iii)
the mixing of the sources into the observations is linear; and (iv),
the number of observations is larger than or equal to the number of
sources.

All of these assumptions hold for our data set: (i) it consists of a
spatially stable mixture of the activities of temporarily independent
quiet Sun and artifactual sources; (ii) the probability densities of
the sources are non-Gaussian; (iii) the superposition of the different
sources arising from quiet Sun, explosive events, and noise, is
linear; and (iv), the number of sources is not larger than the number
of wavelengths sampled. The hypothesis of non-Gaussianity of the
probability densities is based on the current understanding of the
filamentary structure of the Transition Region network, and the bursty
nature of explosive events.

Specifically, let there be \( p \)
signals, \( X(t) = \{ x_1(t), \dots x_p(t) \} \), generated as a sum
of the \( N \) statistical independent components (sources) \( S(t) = \{
s_k(t), \, k=1, \dots , N \} \):
\begin{equation}
	\boldsymbol{X}(t)=\boldsymbol{A} \cdot \boldsymbol{S}(t)
\end{equation}
where \( \boldsymbol{A} \) is the unknown mixing matrix defining the weight of each
source. In particular, let $\boldsymbol{X}(t)$ represent the time evolution of the
line profile emitted at a particular location in the SUMER slit. Thus,
the $ \mathbf{x}_i(t)$ is the time evolution of the line profile at wavelength
$\lambda_i$ at that particular location. 

For the determination of $\boldsymbol{S}(t)$, we use the FastICA
algorithm \citep{Hyvarinen:1999a} with the $\log \cosh ( \cdot )$
~approximation to the negentropy contrast function.

In~\citet{Sarro:2008a} a first version of this method was used. There,
we first removed the long timescales component of the time evolution
by thresholding the appropriate coefficients in a wavelet
decomposition of the original signal. These long time scales were
interpreted as the movement of network elements across the slit which
occurs in timescales much longer than those typical of explosive
events. Then, the signal was decomposed into its independent
components and these components sought for explosive events. This
proved to be a promising technique for the preprocessing of SUMER data
cubes with two main drawbacks: (i) the limitation in the maximal
number of independent sources that could be identified, and (ii) the
impossibility to find a common ordering of the independent sources
found for different slices (positions along the slit) of the data
cube, due to intrinsic ambiguities in the ICA algorithms.

\begin{figure*}[tbh]
\centering
\begin{tikzpicture}[level distance=4mm]
\draw (0,0)  node[text width=1.5cm, text badly centered,inner sep=5pt] (X) {$\mathbf{X}$};
\draw (2,1) node[draw,circle,text width=0.5cm, text badly centered,inner sep=2pt] (icaA) {$\mathbf{A}$};

\draw (2,-1) node[text width=0.5cm,text badly centered,inner sep=2pt] (icaS)  {$\mathbf{S}$};

\draw[->] (X) |- (icaA);
\draw[->] (X) |- (icaS);

\draw (4.5,-1) node[text width=2.25cm,inner sep=2pt] (w) {$CWT_{\mathbf{S}}(a,b;\psi)$};
\draw[->] (icaS) -- (w);

\draw (7.5,-1) node[text width=2.25cm,inner sep=2pt] (tw) {$\widetilde{CWT}_{\mathbf{S}}(a,b;\psi)$} ;
\draw[->] (w) -- (tw) node[coordinate,midway] {}
child[grow=90,-] {node[above] {Thresholding}};

\draw (10.5,-1) node[draw,circle,text width=0.5cm, text badly centered,inner sep=2pt] (icaws) {$\mathbf{\hat{S}}$};
\draw[->] (tw) -- (icaws) node[coordinate,midway] {}
child[grow=90,-] {node[above] {Reconstruction}};

\draw (12.5,0) node[text width=1.5cm, text badly centered,inner sep=5pt] (tx) {$\mathbf{\hat{X}}$};
\draw[->] (icaA) -| (tx);
\draw[->] (icaws) -| (tx);

\draw[dashed,gray!90] (1,-2) -- (1,2);
\draw[dashed,gray!90] (3,-2) -- (3,2);
\draw[dashed,gray!90] (11.5,-2) -- (11.5,2);

\draw (0,-2)  node[text width=1.5cm, text badly centered] {Data set};
\draw (2,-2)  node[text width=1.5cm, text badly centered] {ICA};
\draw (6.5,-2)  node {Wavelet descomposition};
\draw (12.5,-2)  node[text width=1.5cm, text badly centered] {Explosive events};
\draw[snake=brace] (1,2) -- (11.5,2) node[above=5pt,midway] {wICA};
\end{tikzpicture}
\caption{Flow chart of the Wavelet enhanced ICA processing for quiet
  Sun removal}
\label{fig:wICA}
\end{figure*}
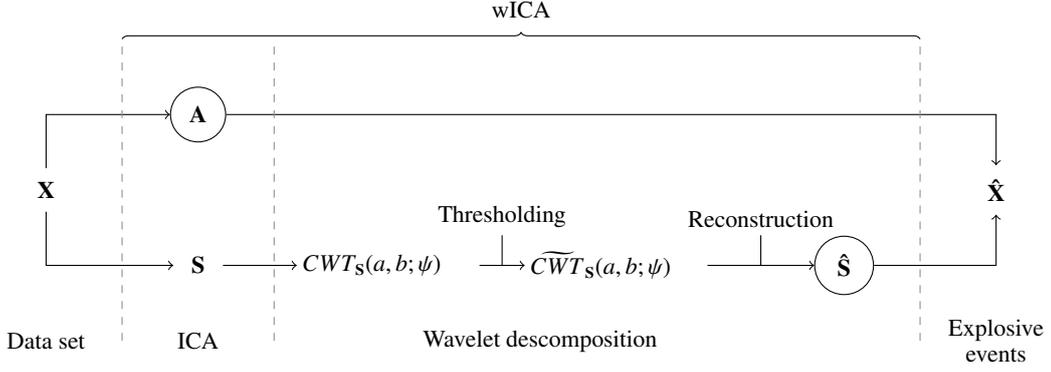

To overcome these limitations, we apply here a reversed version of the
processing scheme presented in~\citet{citeulike:781820}: first we
apply ICA to the data, and then decompose each independent source with
a wavelet analysis. Fig.~\ref{fig:wICA} shows the method in a
schematic way.

When we use ICA to identify independent sources, we often find that
the explosive events appear in independent components that still
contain a significant contribution from typical quiet Sun regions. Let
us model these components as the sum of a high amplitude short lived
component representative of the activity \( s_1(t) \), and a long lived
low amplitude residual quiet signal \( s_2(t) \):

\begin{equation}\label{ec:splitcomponents}
	s_i(t)=s_{1i}(t)+s_{2i}(t),
\end{equation}
where \( s_i \) is the original independent component obtained by the
ICA algorithm. The objective then, is to estimate the quiet Sun
component and subtract it, in order to isolate the explosive event
contribution to the line profile. 

Note that \( s_1 \) has high magnitude and is very localized in time,
while \( s_2 \) has low amplitude and broad band spectrum. We will
use the wavelet decomposition technique to separate these signals,
providing an optimal resolution both in the time and frequency
domains, without requiring the signal to be stationary.

The discrete wavelet transform (DWT) of the independent component \(
s(t) \) reads:
\begin{equation}\label{ec:dwtsignal}
	DWT_s(j,k;\psi)=\int s(t) \psi_{j,k}(y) \, dt ,
\end{equation}
where \( DWT_s(j,k;\psi) \) is the wavelet representation of \( s(t)
\), \( \psi \) is the mother wavelet with \( j \) and \( k \) defining
the time localization and scale,i.e. \( \psi_{j,k}(t)=2^{-j/2}
\psi(2^{-j}t-k)\). Using Eqs. (\ref{ec:splitcomponents}) and
(\ref{ec:dwtsignal}) we can write:
 \begin{equation}
 	DWT_s(j,k;\psi)=DWT_{s_1}(j,k;\psi) + DWT_{s_2}(j,k;\psi) ,
 \end{equation}
where \( DWT_{s_1}(j,k;\psi) \) and \( DWT_{s_2}(j,k;\psi) \) are the wavelet
coefficients obtained by the transformation of the \emph{active} and
\emph{quiet} contributions of the independent component respectively.

In order to subtract the quiet Sun component from the original signal,
we perform a hard-thresholding of the wavelet coefficients.

The resulting set of thresholded wavelet transform coefficients is
inverted, resulting in a denoised version of the original data, with
the quiet Sun component removed. Specifically, we use a hard
thresholding:
\begin{equation} \label{ec:softthresholding}
 \widetilde{DWT}_s(j,k;\psi) = \left\{
\begin{array}{ll} 0 & |DWT_s(j,k;\psi)| \le T_j \\
 \noalign{\medskip}
DWT_s(j,k;\psi) \quad   &  \mbox{ otherwise }
 \end{array}
 \right.
 \end{equation}

The threshold is selected based on the universal model of the noise:
\( T_j = \sqrt{2* \log(n)} \, \sigma\). As \( \sigma \) is typically
unknown, it is estimated based on the median absolute deviation of the
absolute value of the wavelet coefficients, i.e. \( \sigma^2= \mbox{
  MAD} (|DWT_s(j,k;\psi)|)/0.6745 \).

We get separation of the wavelet coefficients into \emph{active} and
\emph{quiet} Sun, i.e. \( DWT_{s_1}(j,k;\psi)=0 \) then \( DWT_{s_2}(j,k;\psi)
\neq 0 \) and viceversa. Then, the wICA-corrected temporal evolution
is:

\begin{equation}\label{ec:wica}
	\hat{\boldsymbol{X}}(t)=\boldsymbol{A} \cdot (s_{11}(t),s_{12}(t),\dots , s_{1N}(t))^T.
\end{equation}

Summarising, the wavelet enhanced ICA (wICA) algorithm can be
described by the following stages:
\begin{enumerate}
\item Apply the conventional ICA decomposition to the time evolution
  of all monochromatic fluxes, thus obtaining the mixing matrix \(
  \boldsymbol{A} \) and \( N \) independent components \( \{ s_1(t),
  s_2(t), \dots , s_N(t) \} \).
\item Calculate \( DWT_{s_i}(j,k;\psi) \) for \( i \in \{1,2, \dots ,
  N\} \)
\item Threshold the wavelet coefficients.
\item Apply the inverse wavelet transform to the thresholded
  coefficients \( \widetilde{DWT}_{s_i}(j,k;\psi) \) to recover only the
  active component \( \{s_{1i}(t)\} \).
\item Signal reconstruction: \( \hat{X} (t)= \boldsymbol{A} \cdot
  (s_{11}(t),s_{12}(t),\dots , s_{1N} (t))^T \).
\end{enumerate}

\begin{figure*}[htp]
   \centering
   \includegraphics[scale=.6]{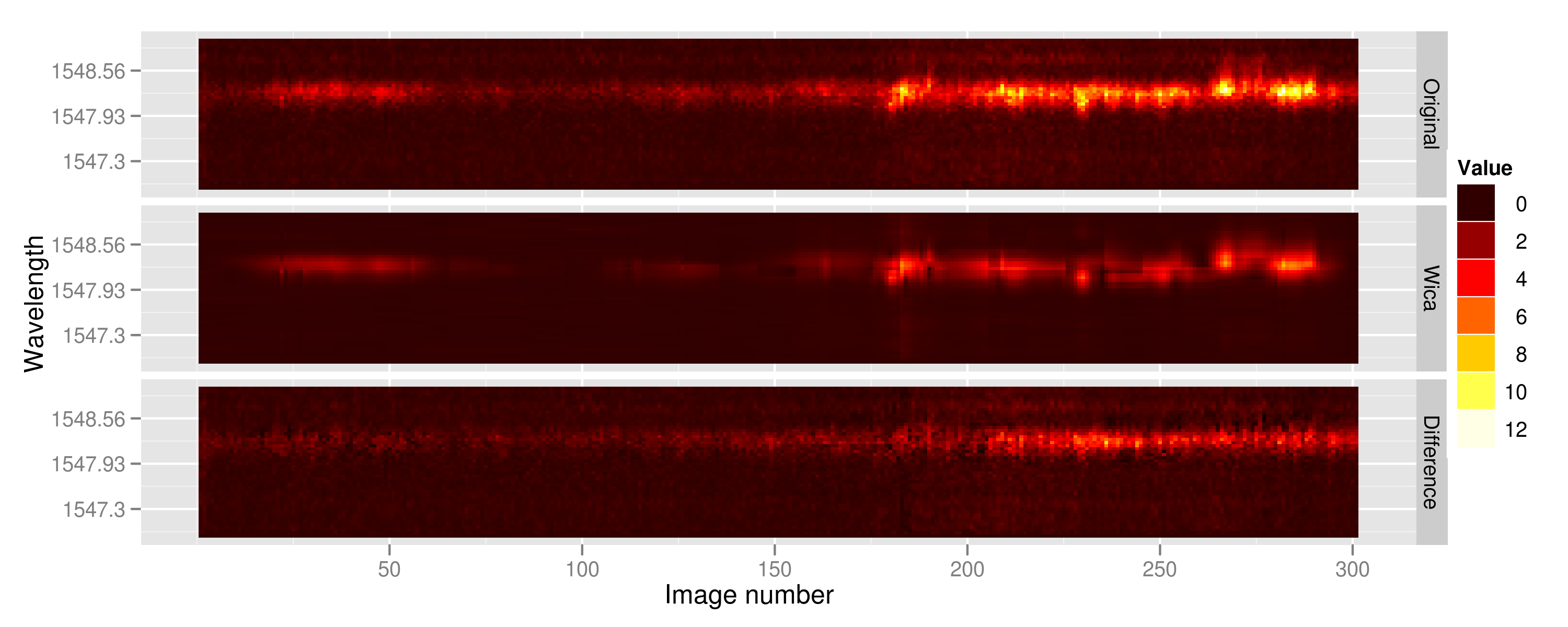}
   \caption{Example of the application of the wICA algorithm to a
     slice of the data cube showing the time evolution of the C~{\sc
       iv} line for a fixed position along the slit. From top to
     bottom we show the original image, the result of the wICA
     algorithm, and the residual image. The images show time in
     minutes along the $x$ axis, and wavelength in Angstroms along the
     $y$ axis.}
   \label{wicaexample}
\end{figure*}

Figure \ref{wicaexample} shows the result of applying the wICA
algorithm to a slice of the first data cube of observations in the
C~{\sc iv} line along the spectral dispersion and time axes. In the
top panel we show the original time evolution of the line profiles for
a fixed latitude; the middle plot shows the result of applying the
wICA filter to the original image, and finally, the bottom panel shows
the residual image.

\section{Results} 
\label{sec:results}

We have applied the techniques described in the previous section to
the dataset obtained with SUMER on May 18th and 19th, 2000. The
preliminary analysis of the results thus obtained comprised the
computation of line radiances, first four line profile moments, and
the line profiles maxima. For the outlier detection stage, only the
four line profile moments were used in order to rank the rasters
according to their non-gaussianity. We intendedly left aside the line
radiances and absolute maximum intesities because we did not want to
bias the statistical analysis of the explosive events towards the
subsample of brightest events. 

The $k$-th order moment ($k \geq 2$) of a given raster is defined as 

\begin{equation}
m_k = \sum_i I_{norm,i} (v_i - m_1)^k 
\end{equation}

where $I_{norm,i}$ is the normalised intensity (normalised dividing by
the integrated emission in the line profile), $v_i$ represents the
Doppler velocity, and

\begin{equation}
m_1 = \sum_i I_{norm,i} v_i 
\end{equation}

is the first order moment or weighted mean of the line profile; the
summation over $i$ is defined only for values above a 33\% of the peak
intensity in order to prevent the noise from contaminating the
estimate.

One of the advantages of the outlier detection method described in the
previous section is that it allows the ranking of line profiles
according to the final weights they are assigned. Drawing a line where
line profiles are no longer explosive events is difficult, and it may
depend on the physcal scenario used to interpret them. We have defined
the samples that we will further analyse as those line profiles with
final weights below 0.9. This is somewhat arbitrary and we do not
claim that all of the line profiles in each sample should
unambiguously be considered as an explosive event. In all cases they
are composed of line profiles with final weights close to zero which
are clearly explosive events under all perspectives, but also of line
profiles that outlie the global distribution for reasons that could be
explained by correlated noise excursions. We have minimized this sort
of ambiguity by imposing thresholds on the signal--to--noise ratio
beyond the typical $3\sigma$ limit for the statistical significance of
the detection.

As mentioned in Sect. \ref{obs}, we find evidence that the slit
position in both observational sequences covers the external regions
of two active regions (NOAA 8998 and NOAA 9004). We have therefore
carried out separate studies for the quiet Sun rasters (latitudes
below -2$^\circ$ in the first sequence and between $\pm5^\circ$ in the
second) and for those potentially affected by the active region
(the complementary regions to those defined as quiet Sun).

In the case study of the C~{\sc iv} line we have further restricted
the slit positions included in the active region definition. We have
detected a systematic shift in the time-averaged line profiles in the
data cube slices corresponding to latitudes below -5$^\circ$. The
shift increases with decreasing latitude and is due to the appearance
of zeros at the short wavelength side of the images. These may affect
up to eight consecutive pixels at the lower end of the slit. We
interpret this as a problem with the IDL subroutine that corrects for
the geometric distorion of the original SUMER image available in the
archive. This problem does not affect the Ne~{\sc viii} line where the
existence of isolated zeroes at either extreme of the wavelength range
does not seem to be correlated with wavelength shifts in the
time-averaged line profiles.

In order to define a wavelength scale from the pixel position along
the wavelength dispersion axis, we have adopted as reference pixel the
one where the maximum emission is attained in the average line profile
over all available latitudes and times in the quiet Sun as defined
above. We subsequently assigned to this reference pixel, the nominal
wavelength of the corresponding line (1548.21~\AA~ for the C~{\sc iv}
line and 770.43; see \cite{1999A&A...346..285D} for a justification of
this value). We are perfectly aware that these values are the rest
wavelengths for these lines, and that these lines are systematically
red shifted in the solar Transition Region. We have prefered to use the
rest wavelengths because, as we shall see later, there are significant
differences in the distribution of first moments in the scatter plots
of the quiet Sun and active regions, and since the aim of this study
was not the accurate determination of the transition region systematic
redshifts, but the statistical characterization of the explosive event
population at small inclinations (i.e. latitudes), we prefered to use
a consistent reference wavelength for all regions and lines, and
discuss the relative differences amongst them.

The use of the average line profile over all quiet Sun rasters to
define the reference pixel for wavelength calibration (instead of the
values recorded in the FITS headers) is due to the obvious
inconsistency between the reference pixel and the line profiles, that
would result in an average line profile red shifted by around 50 km
s$^{-1}$ if the FITS header values were used.

In Sects. \ref{civqs}--\ref{neviiiar} we present the properties of the
explosive events in relation with the population of typical quiet Sun
line profiles, while in Sect. \ref{propee} we analyse potential
correlations between explosive events properties.

\subsection{\label{civqs}The C~{\sc iv} line at 1548.2~\AA~ in the quiet Sun.}

We first apply the outlier detection technique described in Sect.
\ref{sec:methodology} to a dataset comprising the regions in the two
observational series far from the active regions. As an outcome of
this stage we recover a final weight between 0 and 1 assigned to each
raster. We group rasters in two categories: typical profiles and
outliers. We subsequently use the first category to define the
wavelength scale such that the mean quiet Sun line profile is centred
at the nominal wavelength of 1548.21 \AA.

\begin{figure}[htp]
   \centering
   \includegraphics[scale=.40]{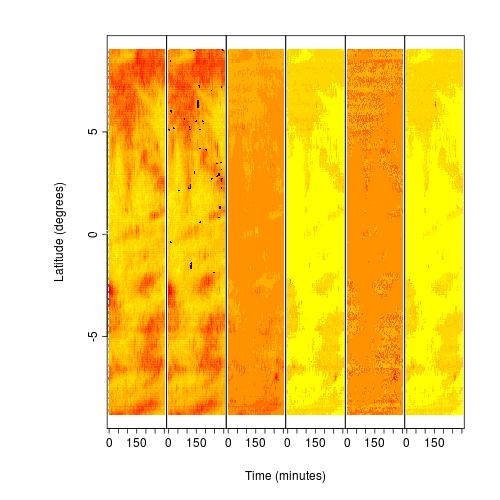}
   \caption{Time-latitude diagram for the first series of
     observations. In the leftmost panel the total line radiance is
     shown in a logarithmic colour scale from yellow (lowest) to red
     (highest values). The next plot to the right shows the same
     diagram with the position of the quiet Sun C~{\sc iv} outliers
     superimposed. Finally, form left to right, the same time-latitude
     diagram with the colour code representing the first, second,
     third, and fourth order moments in a linear scale.}
   \label{time-lat1-cq}
\end{figure}

\begin{figure}[htp]
   \centering
   \includegraphics[scale=.40]{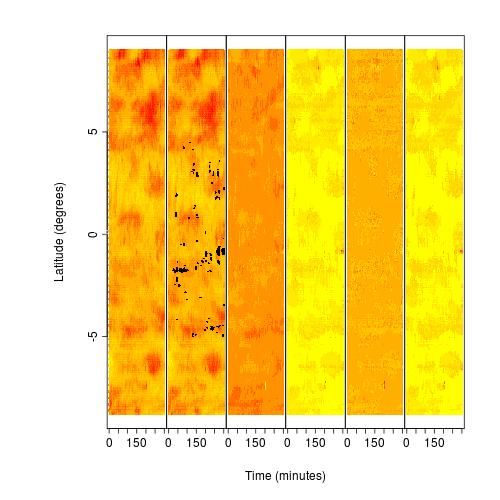}
   \caption{As Fig. \ref{time-lat1-cq} but for the second series of
     observations.}
   \label{time-lat2-cq}
\end{figure}

We show in Figs. \ref{time-lat1-cq} and \ref{time-lat2-cq} the
position of the explosive events in a latitude--time diagram. Since
the SUMER slit position was not corrected for solar rotation, this
diagram is almost equivalent to a latitude--longitude map except for
the non-simultaneity of the measurements. Also, later times in these
figures correspond to western latitudes, so the $x$ axis is reversed
with respect to Figs. \ref{mdi18} and \ref{mdi19}. We have shown in
the leftmost panel the line radiance measured between 1547.7 and
1548.7~\AA in a logarithmic scale. Then, from left to right, the
position of the clear outliers and the first four moments of the line
profiles. The kurtosis (fourth moment) is thus the rightmost panel in
the figure.

Neglecting the variations in the solar rotation with latitude, we can
covert the total time span of the observations (roughly 5 hours) into
an approximate width of 0.17 arcsec/min $\times$ 300 min $\approx$ 50
arcsec. This means that the real observed area is approximately
$300''\times 50''$. The width of each of the images in Figs.
\ref{time-lat1-cq} and \ref{time-lat2-cq} corresponds roughly to this
aspect ratio.

\begin{figure}[htp]
   \centering
   \includegraphics[scale=.40]{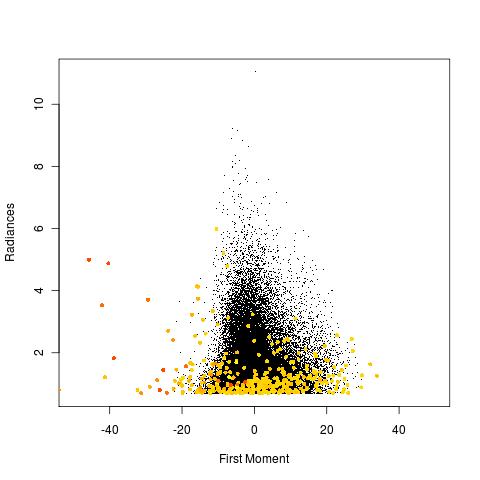}
   \caption{Distribution of first order moments and radiances in the
     quiet Sun zones and the two series of observations. The $x$ axis
     represents the first order moment of the line profile in units of
     km s$^{-1}$, and the $y$ axis, the C~{\sc iv} line radiance in
     units of W m$^{-2}$ sr$^{-1}$. Outliers (explosive events)
       are represented using a colour code indicative of the second
       order moments of the line profiles.}
   \label{civradm1m2}
\end{figure}

Figure \ref{civradm1m2} shows a scatter plot of the first moment of
the line profiles and the integrated line radiances for the
significant detections, defined as line profiles with an integrated
intensity which exceeds the noise level by $7\sigma$, with $\sigma$
the standard deviation of the continuum level between 1546.9 and
1547.2 \AA. This seemingly restrictive threshold is necessary in order
to avoid line profiles that are strongly non-gaussian due to noise
excursions. The common $3\sigma$ threshold is related to the
significance of the detection in the case of a Gaussian signal,
  but not otherwise. This region is affected by the presence of
  several Si~{\sc i} lines and therefore the actual signal-to-noise
  ratio is greater than estimated. We have overplotted the outliers
  using a colour code that indicates the value of the line profile
  second order moment, from yellow (lowest values of the second order
  moment) to red (high values). The scatter plot shows that the vast
  majority of explosive events concentrate in the low range of line
  radiances. There are only a few explosive events with radiances
  above 2 W m$^{-2}$ sr$^{-1}$, and they are all characterized by
  large second order moments and net blue shifts. There are hints of a
  trend in the sense of broader line profiles (redder dots) in the
  negative side of the $m_1$ axis (predominantly blue shifted
  profiles), although first order moments have to be interpreted with
  care: if a line profile has three symmetric components (blue, red
  shifted, and central components) the first moment of the line
  profile can be vanishingly small even if the separate components
  arise at significant velocities.

There is one further aspect to be remarked: the asymmetry in the
distribution of $m_1$ values for line profiles not in the outlier
category. It is due to a separate population of low brightness rasters
shifted to the red typically by 10--15 km s$^{-1}$ with respect to the
median value of $m_1$ in the sample. The corresponding rasters do not
outstand in the distributions of line variance or kurtosis, but are
more frequently characterized by negative asymmetries in the third
order moments. They thus represent line profiles shifted by up to 30
km s$^{-1}$ to the red, with typical line widths, slightly asymmetric,
and they appear as outliers if we lower the threshold for outlier
detection. If the Doppler shifts were due to the Si~{\sc i} 1548.72
\AA\ satellite line, the line profile would be skewed towards longer
wavelengths. This will be interpreted in terms of MHD models of
explosive events in Sect. \ref{discussion}.

The total number of outliers is 372, all of them inside the range of
integrated intensities defined by the non-outliers, mostly in the low
brightness end of the distribution. Therefore, we confirm previous
results in the sense that explosive events are not brighter than the
typical quiet Sun emission.

\begin{figure}[htp]
   \centering
   \includegraphics[scale=.40]{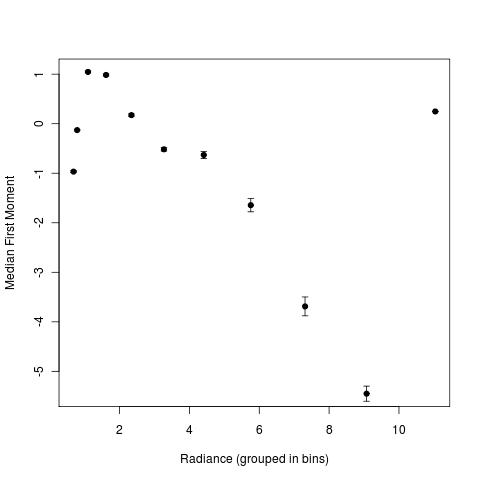}
   \caption{The median wavelength shift as a function of the mid-point
     of radiance bins. The $y$ axis represents the median wavelength
     shift in units of km s$^{-1}$ for several integrated line radiance
     bins.}
   \label{civqsm1radbins}
\end{figure}

Finally, we would like to note that the median wavelength shift is
correlated with the integrated intensity averaged over bins, as shown
in Fig. \ref{civqsm1radbins}. The figure shows the median of the
first moments obtained grouping the data in bins of integrated
intensity. One hundred bootstrap samples are generated in each bin,
and the standard deviation of the medians calculated for all the
bootstrap samples in each bin, shown as error bars. Except in the
first few bins (characterised by small integrated intensities and very
populated) the median first order moment is blue shifted, although it
has to be recalled that the wavelength calibration is done assuming an
average line profile centred at the rest wavelength, and therefore,
the actual values have to be taken with caution. It is only the
relative differences and the actual correlations that are to be
interpreted in the next section. In any case, this proves that the
red shifted, blue skewed profiles discussed in previous paragraphs are
only a small fraction of the total number of line profiles in the low
radiance bins. 

\begin{figure}[htp]
   \centering
   \includegraphics[scale=0.5]{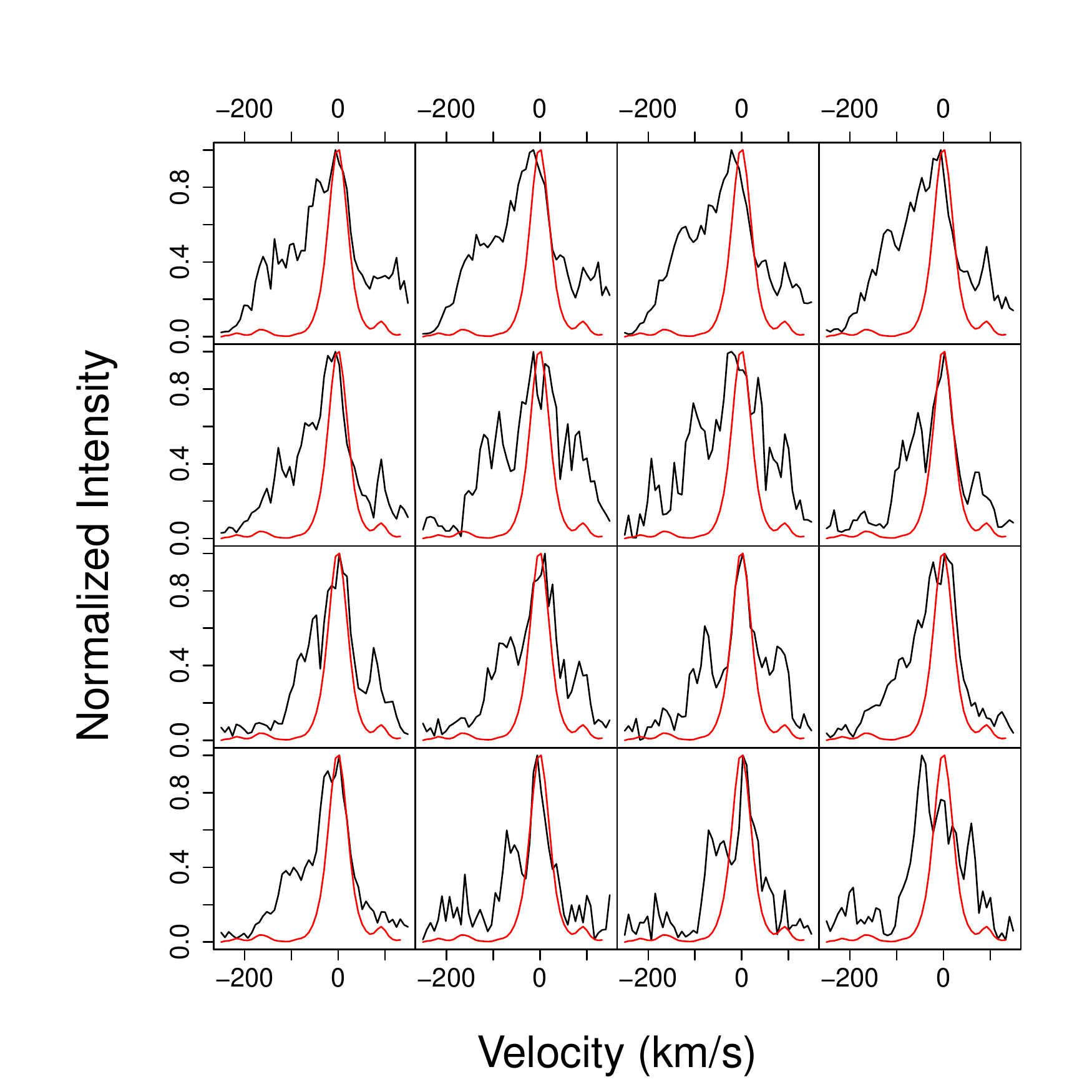}
   \caption{Line profiles in the outlier category of the C~{\sc iv}
     line in the quiet Sun regions. The average quiet Sun profile is
     plot in red.}
   \label{civ.qs.ees}
\end{figure}

We have included in Fig. \ref{civ.qs.ees} plots of the first sixteen
elements in the outlier category, with the average quiet Sun profile
superimposed in red.

\subsection{The C~{\sc iv} line at 1548.2~\AA~ in the active regions.}

\begin{figure}[htp]
   \centering
   \includegraphics[scale=.40]{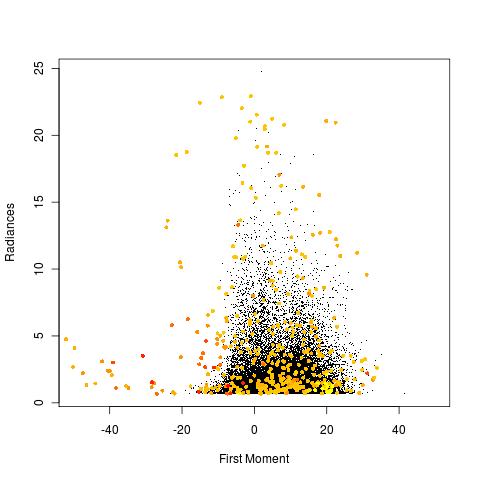}
   \caption{Distribution of first order moments and radiances in the
     proximity of the two active regions NOAA 8998 and NOAA 9004. The
     $x$ axis represents the first order moment of the line profile in
     units of km s$^{-1}$, and the $y$ axis, the C~{\sc iv} integrated
     line radiance in units of W m$^{-2}$ sr$^{-1}$. Outliers
         (explosive events) are represented using a colour code
         indicative of the second order moments of the line profiles.}
   \label{civradm1m2AR}
\end{figure}

Figure \ref{civradm1m2AR} shows the radiance--first moment diagram for
the latitudes potentially affected by the two active regions NOAA 8998
and NOAA 9004. Using the same definition and thresholds for the
outlier detection stage, we find many more extreme events (Doppler
velocities of the order of 200 km s$^{-1}$). Many of these extreme
events appear at small values of the first order moment due to the
line symmetry. 

Compared to Fig. \ref{civradm1m2}, we find that, in the vicinity of
the active regions, the integrated line intensity of both the
explosive events and the population of non outliers can be higher by a
factor 2.5 with respect to the maximum quiet Sun values. It is
important to remark that while the brightest explosive events far from
the active regions were significantly dimmer than the maximum
radiances encountered in the non outlier category (by a factor 0.5),
in the proximities of the active regions the distribution of radiances
of explosive events reaches values as bright as those attained in the
non-outlier sample.

It is also remarkable the increase in relative importance of the red
shifted component amongst the non outliers. While in the quiet Sun we
encounter a main distribution centred around the vanishing first
order moment line as expected, and an additional component of red
shifted line profiles, here the probability density of first order
moments in the low radiance regime peaks around 10 km $s^{-1}$ (it has
to be recalled that, since the definition of the rasters as belonging
to the quiet Sun or active regions is independent of the observational
setup, we have used the same reference pixel and dispersion relation
in the wavelength calibration of the two sets). Although this effect
may be related to the already known difference between the transition
region average redshift in the quiet Sun and active regions \citep[see
  e.g.][]{1999A&A...349..636T}, the magnitude of the difference is
larger in our dataset than previously reported.

\begin{figure}[htp]
   \centering
   \includegraphics[scale=.40]{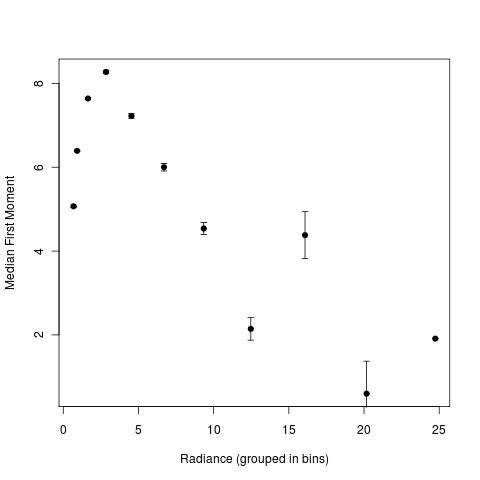}
   \caption{The median wavelength shift in the proximity of the two
     active regions NOAA 8998 and NOAA 9004, as a function of the
     mid-point of radiance bins. The $y$ axis represents the median
     wavelength shift in units of km s$^{-1}$ for several integrated
     line radiance bins.}
   \label{civarm1radbins}
\end{figure}

Figure \ref{civarm1radbins} shows a similar behaviour in the
integrated line radiance as that described in the previous section for
the quiet Sun, except for the fact that in the outskirts of the active
regions, the correlation/anticorrelation always appears at net red
shifts. The behaviour in the last bins is due to the small sample
sizes.

\begin{figure}[htp]
   \centering
   \includegraphics[scale=0.5]{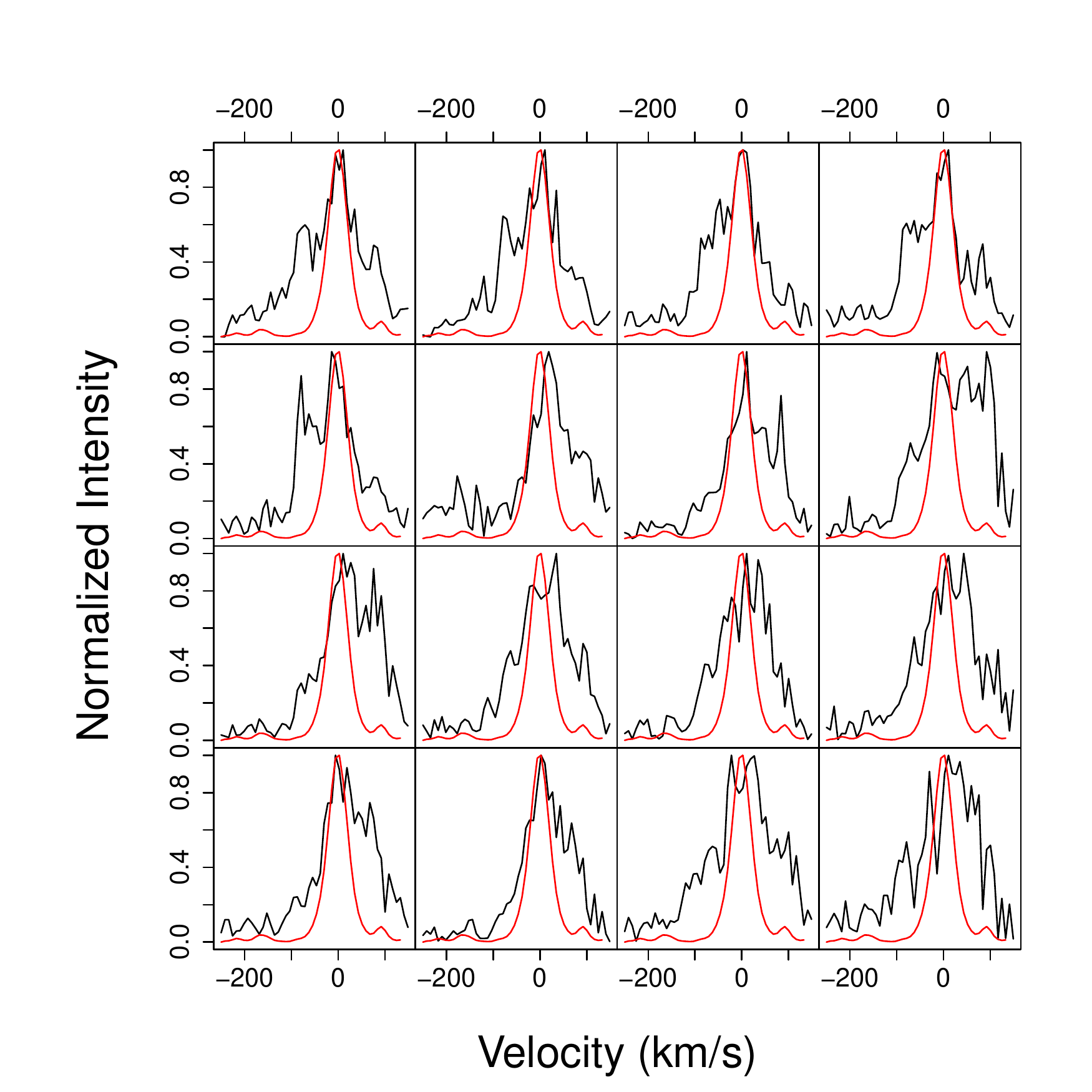}
   \caption{Line profiles in the outlier category of the C~{\sc iv}
     line in the proximity of the active regions. The average quiet
     Sun profile is plot in red.}
   \label{civ.ar.ees}
\end{figure}

We have included in Fig. \ref{civ.ar.ees} plots of the first sixteen
elements in the outlier category for the latitudes affected by the
active regions.

\subsection{The Ne~{\sc viii} line (770.4~\AA) line in the quiet Sun.}

\begin{figure}[htp]
   \centering
   \includegraphics[scale=.40]{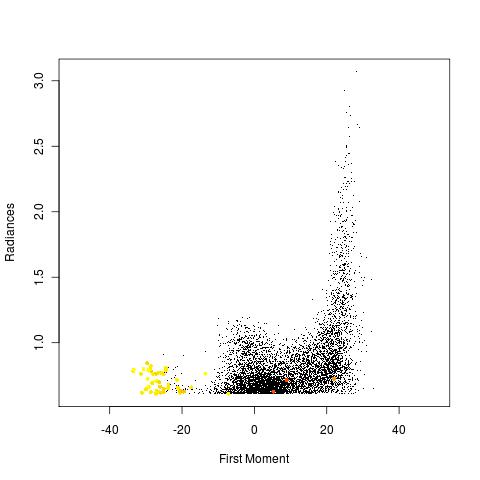}
   \caption{Distribution of first order moments and radiances of the
     Ne~{\sc viii} line in the quiet Sun zones. The $x$ axis
     represents the first order moment of the line profile in units of
     km s$^{-1}$, and the $y$ axis, the Ne~{\sc viii} integrated line
     radiance in units of W m$^{-2}$ sr$^{-1}$. Outliers (explosive
     events) are represented using a colour code indicative of the
     second order moments of the line profiles.}
   \label{neviiiradm1m2}
\end{figure}

\begin{figure}[htp]
  \centering
  \includegraphics[scale=.40]{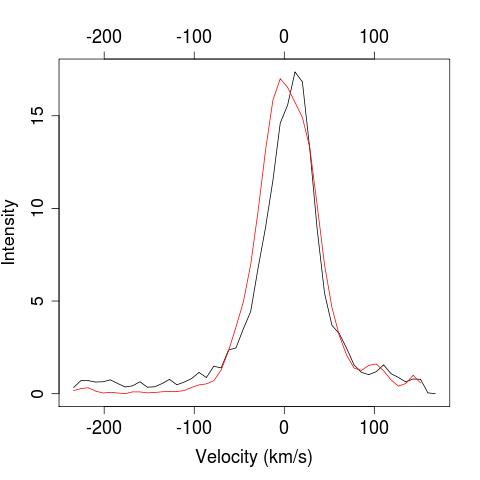}
  \caption{One of the brightest Ne~{\sc viii} line profiles at the tip
    of the brach of red shifted line profiles in Fig.
    \ref{neviiiradm1m2} corresponding to the quiet Sun regions
    (black). The red line corresponds to the mean line profile in
    these quiet Sun rasters. The $y$ axis shows the intensity in the
    Ne~{\sc viii} line in units of W m$^{-2}$ sr$^{-1}$ \AA$^{-1}$. }
  \label{neviii.bright.red}
\end{figure}

The Ne~{\sc viii} line at 770.4~\AA~appears in second order at a
wavelength of 1540.85 \AA. Figure \ref{neviiiradm1m2} shows three
populations in the integrated line radiance and first order momentum
diagram: the bulk of the scatter plot is composed of line profiles
centered around the assumed laboratory wavelength as expected (first
population); there is also a clear branch of rasters that show a
Doppler shift of 20 km s$^{-1}$ and radiances greater than or equal to
the maximum radiances in the bulk cluster (second population; see
Fig. \ref{neviii.bright.red} for an example of the brightest
profiles in this category); finally, there seems to be evidence for a
much smaller cluster of rasters characterized by low radiances and
blueshifts between 20 and 40 km s$^{-1}$ (third population, shown in
colour in Fig. \ref{neviiiradm1m2}).

The second population is characterised by asymmetric line profiles
with enhanced blue wings while the third population is composed of 70
rasters with blue shifts equivalent to Doppler velocities greater than
20 km s$^{-1}$. Visual inspection of the line profiles in this third
category shows that they do not correspond to explosive events. There
is no correlation with latitude but they are concentrated in time in
the first 30 minutes of the first dataset. The line profiles are
asymmetric, with enhanced red wings (as opposed to the second
population line profiles). Since their appearance is concentrated in
groups of two to four consecutive latitudes and intermittent in time,
and they do not show atypical line profile widths, asymmetries or
shapes, we interpret them as instrumental shifts occuring at the
initial phase of the observations. The outliers of the distribution
according to the PCOut algorithm are limited to a number of 47,
and they all fall in this third category of blue shifted line profiles
.

\subsection{\label{neviiiar}The Ne~{\sc viii} line at 770.4~\AA~ in the active regions.}

\begin{figure}[htp]
   \centering
   \includegraphics[scale=.40]{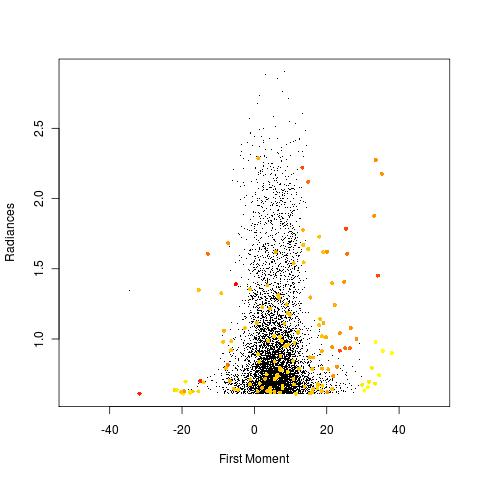}
   \caption{Distribution of first order moments and radiances of the
     Ne~{\sc viii} line in the proximity of the two active regions
     NOAA 8998 and NOAA 9004. The $x$ axis represents the first order
     moment of the line profile in units of km s$^{-1}$, and the $y$
     axis, the Ne~{\sc viii} integrated line radiance in units of W
     m$^{-2}$ sr$^{-1}$. Outliers (explosive events) are
       represented using a colour code indicative of the second order
       moments of the line profiles.}
   \label{neviiiradm1m2AR}
\end{figure}

Figure \ref{neviiiradm1m2AR} shows the radiance--first moment diagram
for the Ne~{\sc viii} line at 770.4~\AA~in the vicinity of the two
active regions NOAA 8998 and NOAA 9004. The diagram is different from
the equivalent plot in the quiet regions around the equator, in that
the main bulk of rasters is no longer centred around the $m_1=0$
vertical line, but around $m_1\approx 7$km s$^{-1}$, as judged from the
histogram.
 
While the quiet Sun outlier analysis of the Ne~{\sc viii} line at
770.4~\AA~ did not yield a significant sample of explosive events but
only a few tens of blue shifted lines, the rasters in the vicinity of
the active regions show significant activity in the form of explosive
events. If we examine the sample of outliers, we clearly recognise an
imbalance between red and blue asymmetries. The most conspicuous
explosive events in the C~{\sc iv} line in quiet Sun and active
regions are characterised by predominant blue shifts in their first
moments. Explosive events in the Ne~{\sc viii} line, on the contrary,
show predominantly red shifted values of $m_1$ as shown in Fig.
\ref{neviiiradm1m2AR}. The total number of outliers is 144
(significantly less than the total number of outliers in the C~{\sc
  iv} outlier category) and although many of them are undoubtedly
explosive events, some can be interpreted as line profiles
non-gaussian due to noise fluctuations.

\begin{figure}[htp]
   \centering
   \includegraphics[scale=0.5]{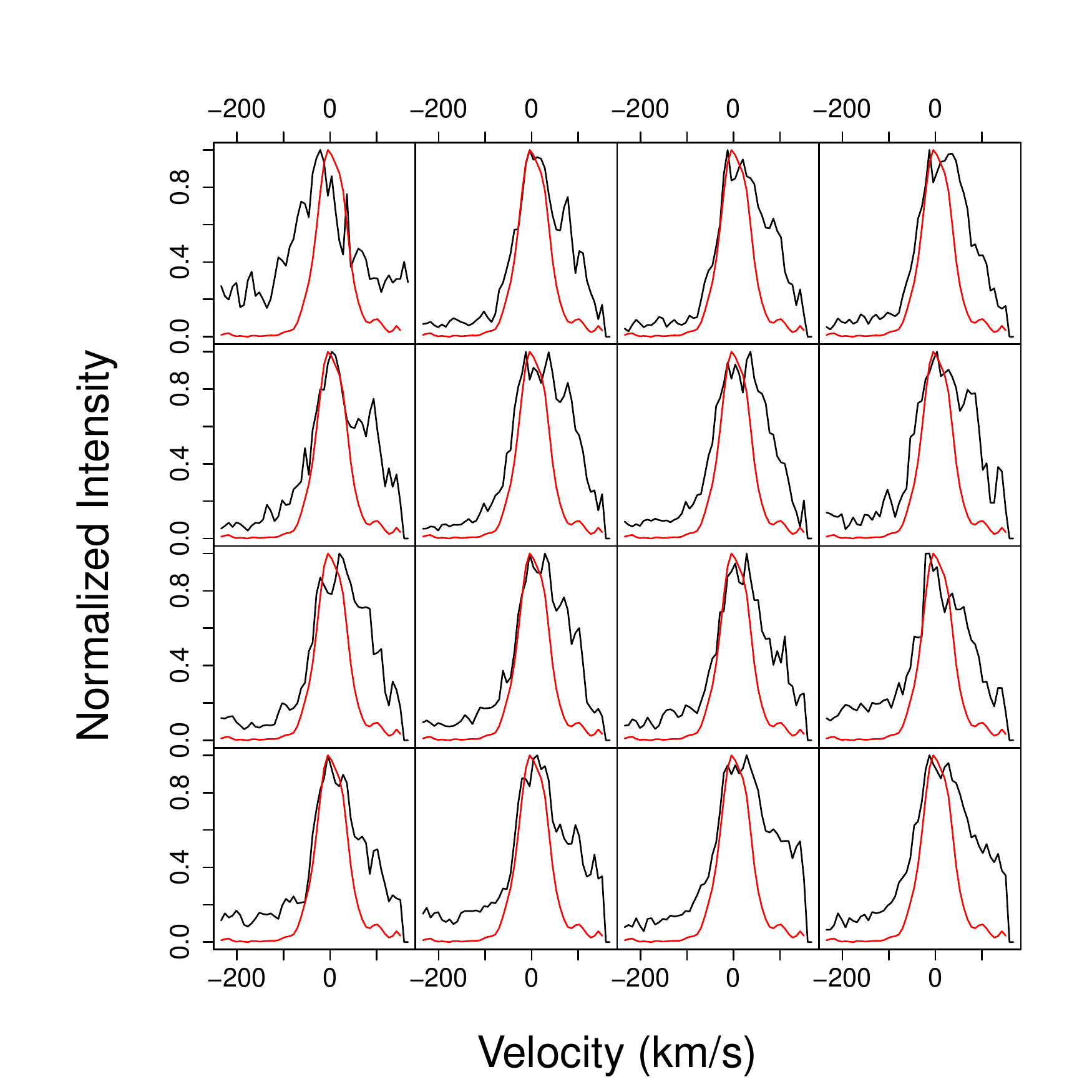}
   \caption{Line profiles in the outlier category of the Ne~{\sc viii}
     line in the proximity of the active regions. The average quiet
     Sun profile is plot in red.}
   \label{neviii.ar.ees}
\end{figure}

We have included in Fig. \ref{neviii.ar.ees} plots of the first
sixteen elements in the Ne~{\sc viii} line outlier category for the
latitudes affected by the active regions.

\subsection{\label{propee} Properties of the red and blue wings of explosive events}

In the following, we analyse and list the properties of the red and
blue wings of explosive events line profiles in these new larger
samples obtained with the methodology described in Sect.
\ref{sec:methodology}. We hope that this summary of the statistical
properties of the new samples will help guide future numerical
simulations of explosive events, by pointing to the observational
trends, or absence of correlations that need be reproduced or
explained by them.

We have defined the red (blue) wing as the line profile to the right
(left) of the rest wavelength (as defined above) in the wICA denoised
spectral images. The red edge of the spectral images correspond to a
Doppler red shift of approximately 150 km s$^{-1}$, while the blue
edge blue shift is sensibly larger ($\approx$ -250 km s$^{-1}$) due to
the rest wavelength not being centred in the window. In the following,
we study the red and blue spectral line wings in two windows 150 km
s$^{-1}$ wide at each side of the rest wavelength. For each window we
compute the line radiance ($F_{red}$, $F_{blue}$), the first and
second line moments ($m_{1,red}$, $m_{1,blue}$, m$_{2,red}$, and
$m_{2,blue}$), and the velocity at which the 20\% of the maximum line
intensity is first attained. These latter values (one for each wing)
are taken as indicative of the maximum Doppler shifts. We believe that
the characterisation of explosive events involves all of these
parameters. Apparently, the second order moments carry similar
information to the velocities at the 20\% of the maximum flux, but
careful visual examination of the explosive events shows that this is
not the case, and we can find very different line profiles with
roughly equal second order moments. In some cases the line profile can
be decomposed into several resolved components, but never is a
gaussian fit an adequate description for them.

\begin{figure*}[htp]
   \centering
   \includegraphics[scale=.3]{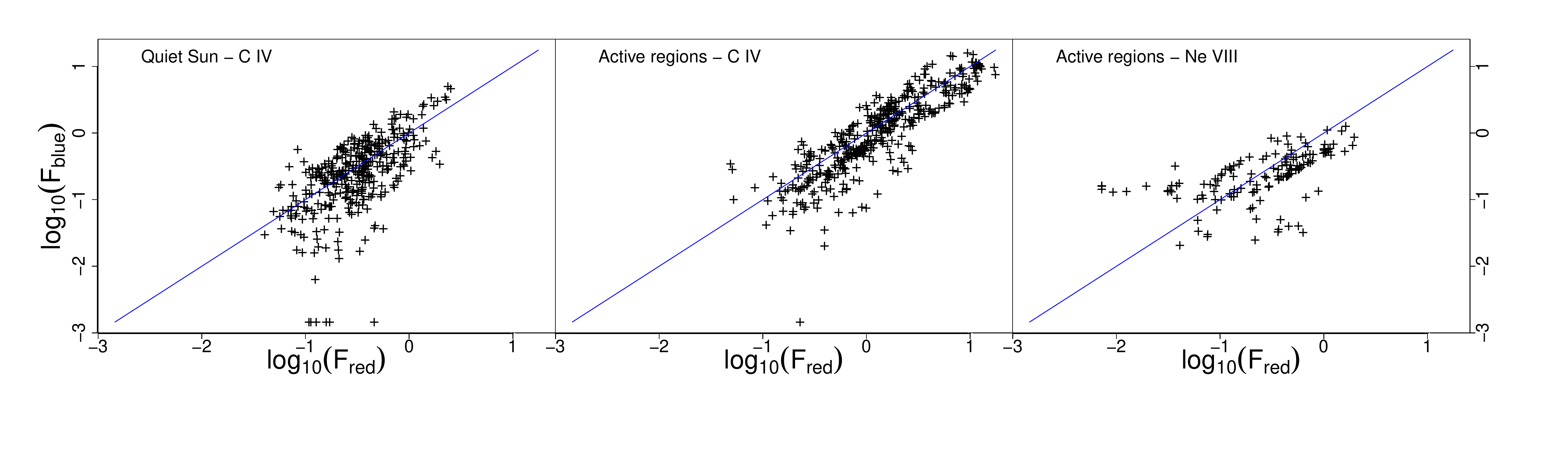}
   \caption{Scatter plots of the radiance emitted in the red and blue
     wings ($x$ and $y$ axes respectively) of explosive events in the
     quiet Sun C{\sc iv} line profiles (left panel), and near the
     active regions in C~{\sc iv} (middle) and Ne~{\sc viii} (right).}
   \label{redbluerads}
\end{figure*}

\begin{figure*}[htp]
   \centering 
   \includegraphics[scale=.3]{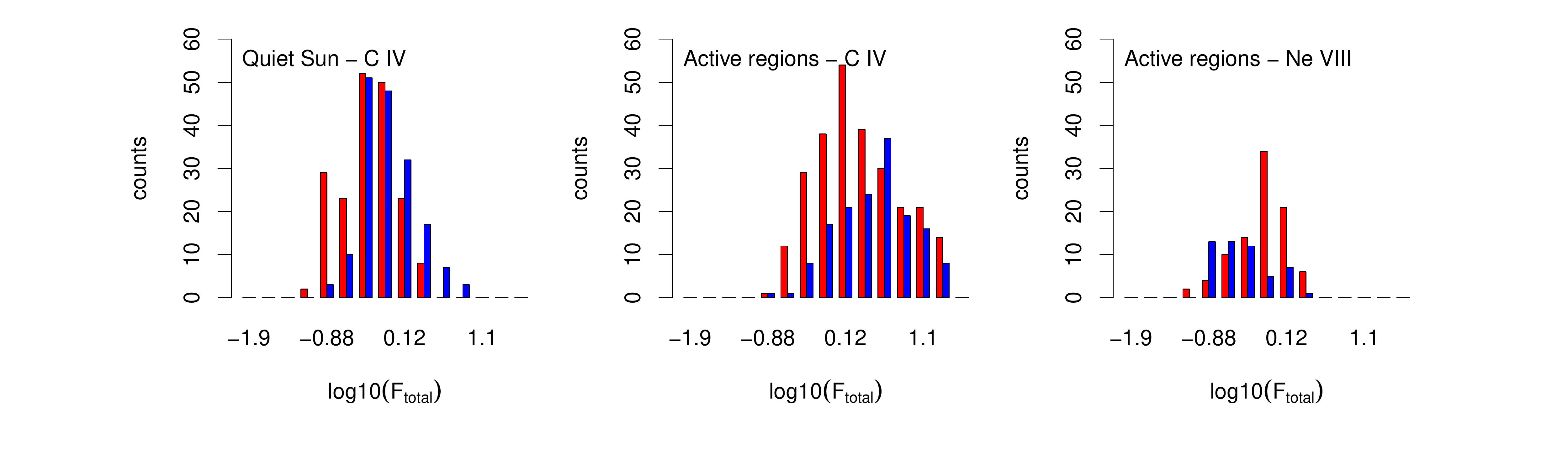}
   \caption{Histograms of the number of explosive events in bins of
     the total line radiance. Red bars correspond to explosive events
     with red wings brighter than the blue wings, and blue bars, to
     blue wings brighter than the red ones. The left and central
     panels correpond to the C~{\sc iv} line in the quiet and active
     regions respectively, and the rightmost panel shows the same
     histogram for the Ne~{\sc viii} line in the vicinity of the
     active regions. }
   \label{radhists}
\end{figure*}

First, we concentrate in the distribution of radiances in the samples.
Figures \ref{redbluerads} and \ref{radhists} show the distribution of
radiances in the red and blue windows and in the three study cases
with significant amounts of explosive events, i.e., the quiet Sun
regions in the C~{\sc iv} line, and the regions in the proximities of
active regions NOAA 8998 and NOAA 9004 in C~{\sc iv} and Ne~{\sc
  viii}. The scatter plots and histograms show hints that the quiet
Sun explosive events in the C~{\sc iv} line are composed of two
populations: one with brighter red wings which is more numerous in the
fainter end of the distribution of total radiances) and one with blue
wings brighter which dominates the population at the bright end of the
distribution. In the proximities of the active regions, explosive
events with brighter red wings still dominate the faint end of the
distribution, but the situation in the bright end is
balanced. Finally, the situation in the Ne~{\sc vii} line is
different, with predominantly red explosive events dominating both at
the faint and bright ends, and predominantly blue events being more
frequent at intermediate radiances.  

\begin{figure*}[htp]
   \centering
   \includegraphics[scale=.3]{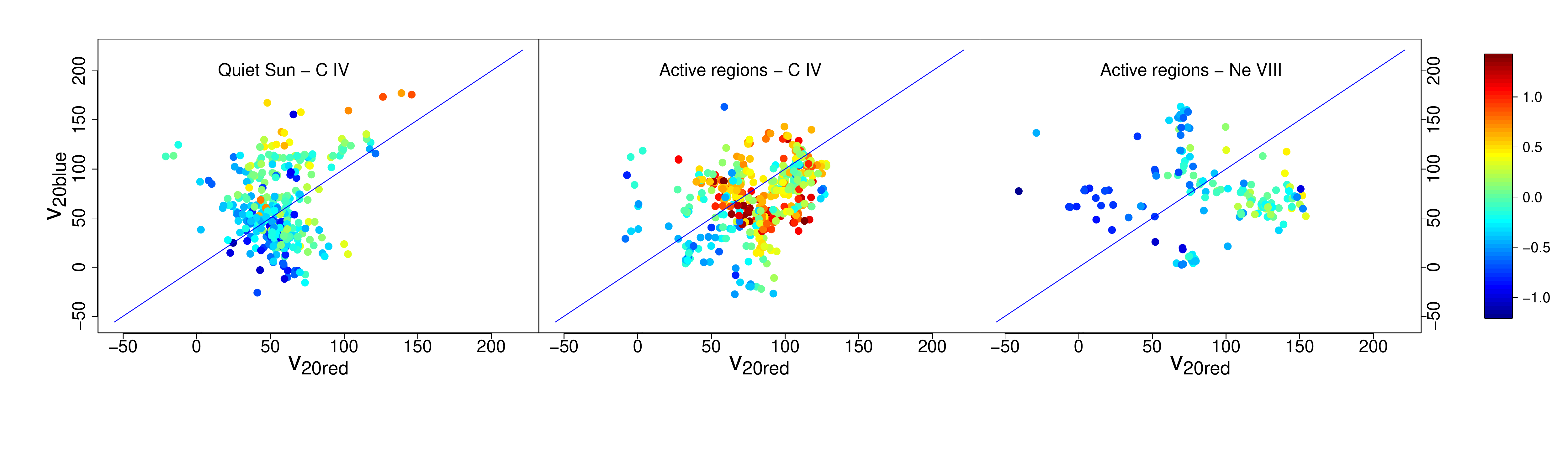}
   \caption{Scatter plots of the velocities at which the 20\% of the
     maximum line profile is attained at each side of the maximum
     emission ($x$- and $y$- axes respectively). The left and central
     plots correspond to the C~{\sc iv} line in the quiet Sun regions
     and in the proximities of the active regions; the rightmost
     scatter plot corresponds to the Ne~{\sc viii} line in the
     proximities of the active regions. The colour code represents the
     total flux radiance in a logarithmic scale. Blue correponds to
     the lowest values of the line radiance while red corresponds to
     the highest values.}
   \label{20quantiles}
\end{figure*}

Next, we concentrate in the distribution of maximum velocities and
their correlation with the total line radiance. Figure
\ref{20quantiles} represents the Doppler shift at which the line
profile decays down to the 20\% of the maximum in the red and blue
wings (the colour code reflects the total line radiance in a
logarithmic scale). This is used as an indicator of the maximum
velocities that characterise the explosive event.  In the quiet Sun
regions (leftmost plot), the scatter plot shows a tendency for the
C~{\sc iv} lines to have larger maximum blue shifts than maximum red
shifts (explosive events above the diagonal) except in the region of
small velocities (below 75 km s$^{-1}$) where maximum red shifts
larger than the maximum blue shifts are more frequent (a situation
similar to that found in Fig. \ref{radhists} for the wing
brightness). Brighter events tend to show larger maximum velocities as
expected, and the brightest events in this category have maximum
velocities which are larger in the blue window.

In the scatter plot corresponding to the C~{\sc iv} line in the
proximities of the active regions, larger maximum red shifts are
systematically more numerous, and there seems to be a separation at
$v_{20}\approx$60--70 km s$^{-1}$.  Explosive events with maximum
velocities larger than these values are brighter and less asymmetric
on average. It is interesting to note that the two scatter plots
showing explosive events in the C~{\sc iv} line are complementary,
with quiet Sun explosive events densely populating the lower left
corner of the diagram and active region explosive events occupying
preferentially the adjacent region to the upper right. 

In the Ne~{\sc viii} line profiles in the outskirts of the active
regions, explosive events show a butterfly diagram with explosive
events characterized by maximum red shifts larger than the maximum
blue shifts being both brighter and slightly more numerous than the
opposite. Line profiles that are totally contained in one of the two
windows only occur in the faint end of the radiance distribution, and
appear in these plots with negative values in one of the axes.

\begin{figure*}[htp]
   \centering
   \includegraphics[scale=.3]{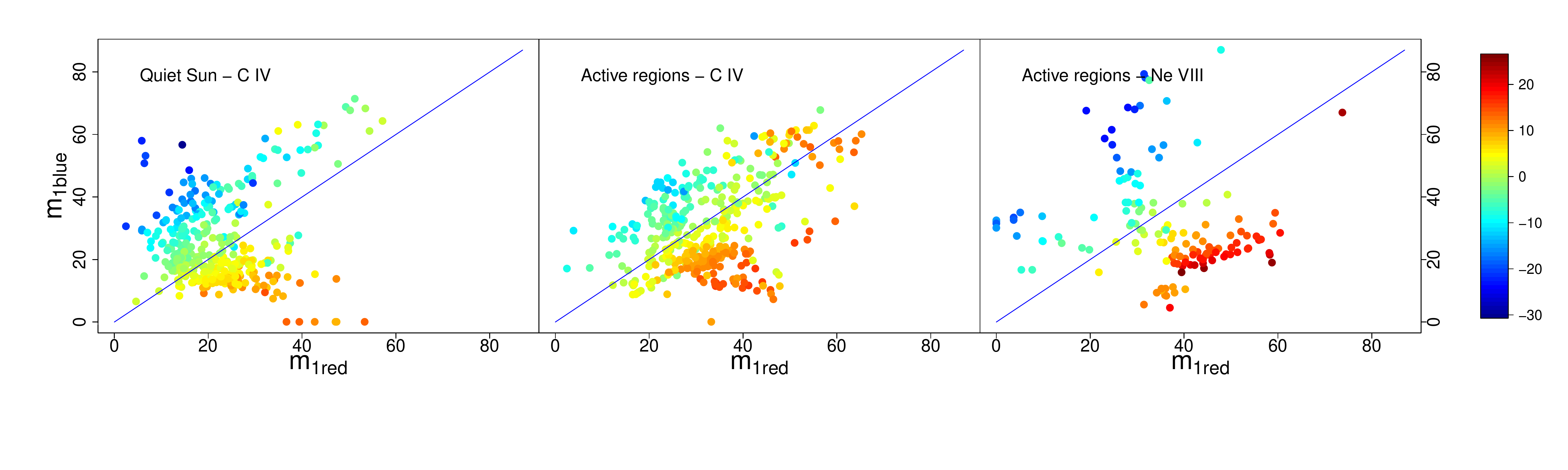}
   \caption{First moments ($m_1$) of the red and blue windows at each
     side of the rest wavelength ($x$ and $y$ axes respectively). The
     colour code represents the second order moment of the red wing
     minus the second order moment of the blue one. Blue correponds to
     the lowest values of the difference while red corresponds to the
     highest values. The plots show, from left to right, the C~{\sc
       iv} line in the quiet Sun and the C~{\sc iv} and Ne~{\sc viii}
     lines in the vicinity of the active regions.}
   \label{firstmoments}
\end{figure*}

Finally in Fig. \ref{firstmoments} we analyse the distribution of
first order moments in relation with the line variance (second order
moment). The $x$ and $y$ axes show the values of the first order
moment of the red and blue wings of the line profile (as defined
above), and the colour code represents the value of the difference
between the second order moments $d_2=m_{2,red}-m_{2,blue}$. Figure
\ref{firstmoments} shows a remarkable difference with respect to
Fig. \ref{20quantiles}. While the Ne~{\sc viii} and quiet Sun C~{\sc
  iv} distributions show basically the same structure in the two
figures, the distribution of C~{\sc iv} explosive events in the
proximities of the active regions shows that profiles that appear with
small first order moments (lower left corner in the middle panel of
Fig. \ref{firstmoments}) are characterised by large maximum velocities
$v_{20}$, and thus, appear at the upper right corner in the middle
panel of Fig. \ref{20quantiles}.

\begin{figure}[htp]
   \centering
   \includegraphics[scale=0.5]{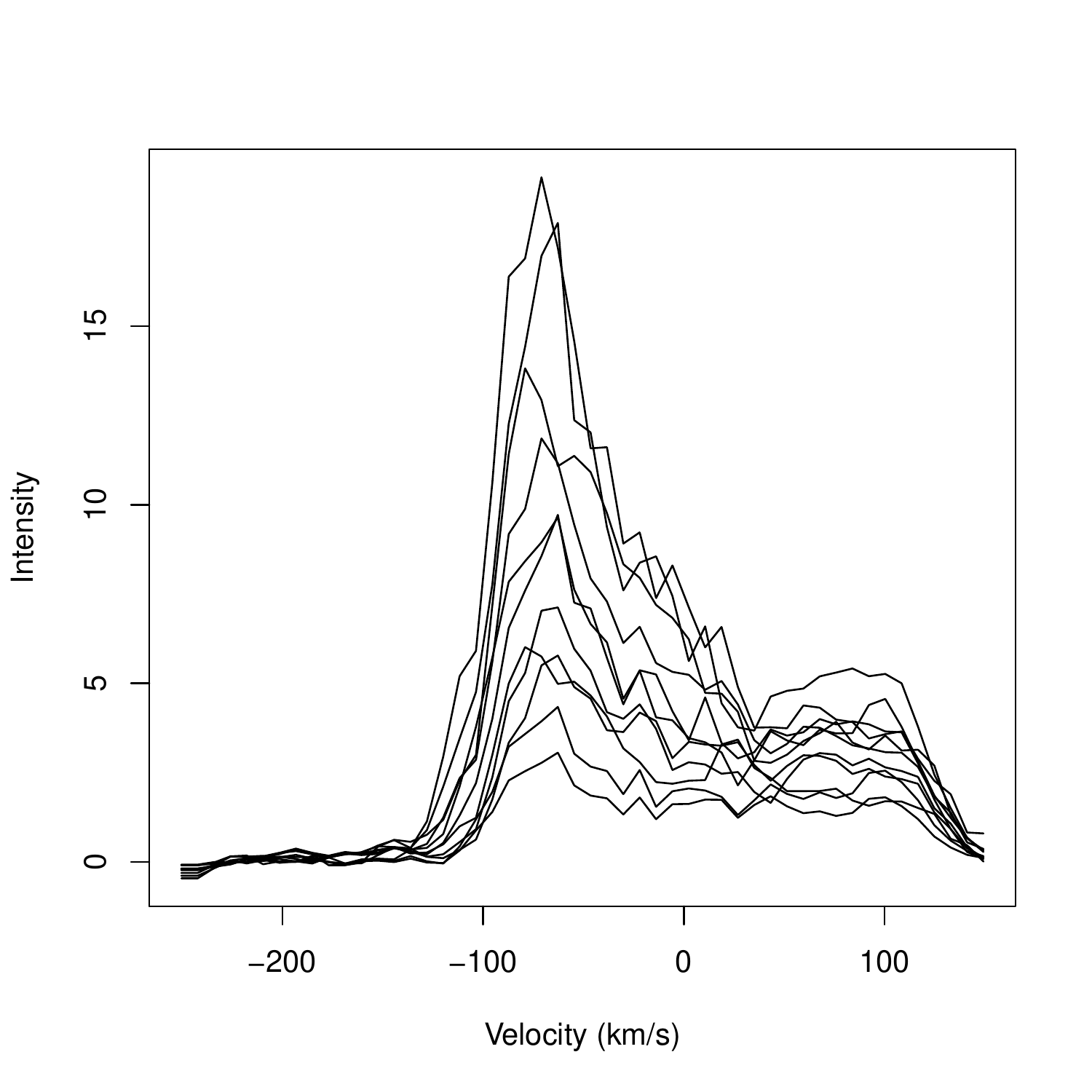}
   \caption{Atypical examples of C~{\sc iv} explosive events in the
     proximities of the active regions NOAA 8998 and NOAA 9004. The
     $y$ axis shows the line intensity in units of W m$^{-2}$
     sr$^{-1}$ \AA$^{-1}$. }
   \label{blueees}
\end{figure}

In particular, it is worth noting the presence of several explosive
events characterized by large positive values of $d_2$, $m_{1,red}$,
and $m_{1,blue}$ in the C~{\sc iv}--active regions panel of Fig.
\ref{firstmoments}, breaking the general trend that $d_2$ isarithms
run parallel to the $m_{1,red}=m_{1,blue}$ line. They correspond to
explosive events with one distinct emission component in the blue wing
(and thus a narrow main velocity component), and extended emission in
the red wing that we interpret as arising from plasma with a wide
distribution of velocities which actually extends beyond the maximum
velocity in the blue window. We have included in Fig. \ref{blueees}
a few examples of this category with the aim of helping in the
interpretation of the $m_1$--$m_2$ diagrams.



\section{Discussion\label{discussion}}

The first result that we would like to discuss is the fact that the
red wings are neither sistematically brighter than the blue
ones, nor slower. This is nevertheless a common prediction from the
simulations which is explained in \cite{2009ApJ...702....1H}: red
wings arise from plasma streams travelling in a medium denser than the
medium encountered by the plasma responsible for the blue wing
emission, and thus, since they propagate at the Afvén speed, they are
characterized by smaller velocities. This is expected in stratified
media, with density decreasing towards the observer. Furthermore and
for the same reason, the plasma receding from the observer is
compressed leading to enhanced emission.

\begin{figure*}[htp]
   \centering \includegraphics[scale=.3]{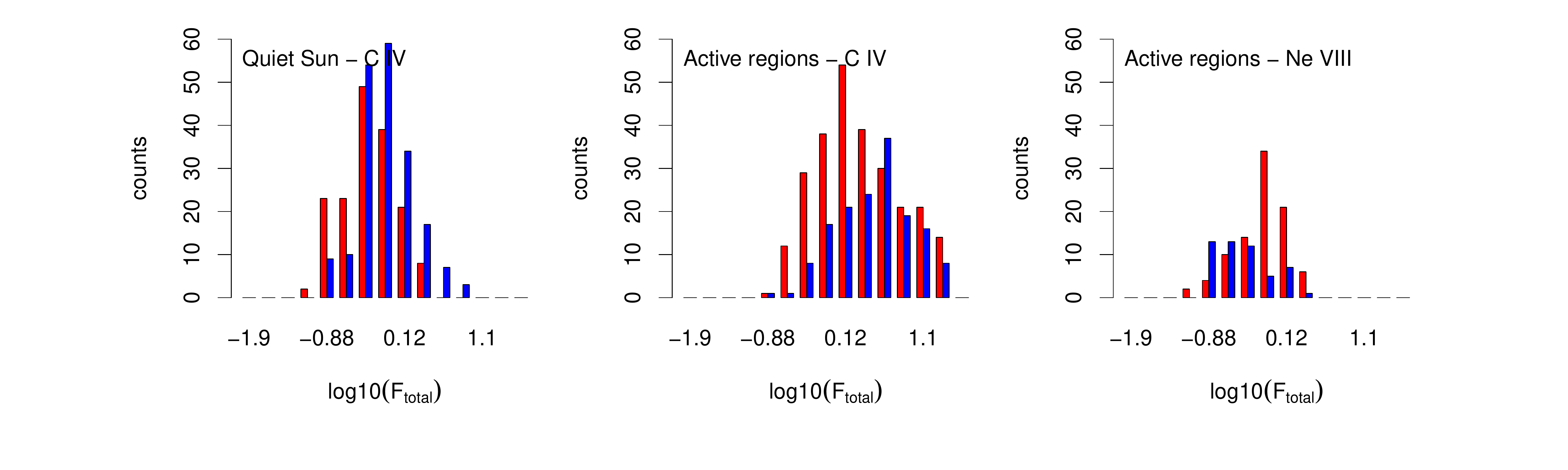}
   \caption{Histograms of the number of explosive events in bins of
     total line radiance. Similar to Fig. 15, but here for velocity
     thresholds: red bars correspond to explosive events with maximum
     velocities in the red wing larger than the corresponding blue
     maximum velocity, and blue bars to larger maximum velocities in
     the blue.
   }
   \label{v20hists}
\end{figure*}

The problem arises in the interpretation of the low radiance explosive
events in the quiet Sun rasters: their red wings are brighter than the
blue ones as expected, but their maximum Doppler shifts in the red
wing are also larger than the maximum blue shifts, as shown in
Figs. \ref{20quantiles} and \ref{v20hists}. The opposite also applies:
the brightest events have larger maximum Doppler shifts in the blue
wing as expected, but also more flux in the blue wings, and this is
also not predicted by the simulations.

Our interpretation of the results shown in the previous sections is a
complex scenario where explosive event properties are defined by
several elements. The most important one is the magnetic flux
available for reconnection. This is evident from the comparison of the
characteristics of the explosive events in the quiet Sun and near the
active regions in Fig. \ref{20quantiles}, that the properties of the
line profiles are almost complementary, to the extent that the small
overlap of explosive events near the active regions showing
characteristics of low energy events and viceversa can be explained as
due to the coarseness of the definition of quiet Sun and active
region. The fact that the available magnetic flux is an important
ingredient in explosive events is not new: the fact that its influence
dominates over that of others is.

The analysis of the next important element that defines the properties
of explosive events is, unfortunately, beyond the present
observational capabilities: it is the small scale geometry of the
magnetic field that dictates the directions of the plasma jets. As
shown by \cite{2009ApJ...702....1H}, the picture of two aligned jets
can only be applied to the reconnection site, and different moving
angles can occur after the jets have interacted with the surrounding
medium. Furthermore, the reconnection geometry varies with time
depending on the phase of the driving mechanism. 


If different deflection angles for the two reconnection jets were
indeed the reason for the unexpected larger maximum velocities in the
red wing, it would imply that the emission detected as explosive
events comes mainly, not from the reconnection site, but from regions
where interaction with the surrounding medium has already changed the
direction of the flow. More simulations with different initial
magnetic field configurations are needed in order to explore this
hypothesis.

The height above the solar photosphere where magnetic reconnection
occurs is also another important ingredient. The effect of a height
variation is twofold. Down in the chromosphere, the expected Alfven
velocity is smaller, and emission from both the reconnection site and
the plasma flows receding from it will appear combined in the same
spectrograph element for a longer time. But also, densities in the
chromosphere are higher than at the base of the Transition Region and
therefore higher radiances should be expected. We have not found
evidence for this correlation between larger line radiances and
smaller maximum velocities, and therefore we must conclude that either
the explosive events cannot be triggered in a wide range of heights,
or the correlation is masked by other effects. 



In general, we find that the brightest C~{\sc iv} explosive events are
also characterised on average by larger maximum velocities. If the
enhanced line radiances were due to a chromospheric origin, we would
expect small maximum velocities, and since this is not the case, we
interpret them as due to a larger amount of magnetic energy ready to
be tranformed into kinetic energy and heating. But we find marginal
evidence in the middle panel of Fig. \ref{20quantiles} that the
brightest events, even if the occupy the region of large maximum
velocities, tend to cluster around $v_{max}\approx$70 km s$^{-1}$.
This could be an indication that the two mechanisms are taking place:
near the active regions there is more magnetic energy available and
thus, explosive events are brighter, but at the same time they occur
preferentially at lower heights in the chromosphere and, as a
consequence, they are bright but relatively slow compared to events
with the same available energy but triggered higher in the
atmosphere. 



Finally, we are tempted to interpret the population of gaussian red
shifted profiles in the quiet Sun in C~{\sc iv} (see Fig.
\ref{civradm1m2}) as a population of low energy explosive events
occuring deep in the chromosphere. But the lack of blue wing
enhancements at all possible line radiances argues against this
hypothesis.

\section{Conclusions}

In this work we have presented a totally automated processing of
homogeneous series of spectral images taken by the SUMER spectrograph
on board SOHO. As a result, we have several samples of explosive
events in the two lines of C~{\sc iv} and Ne~{\sc viii}, in regions of
the quiet Sun and in the outer parts of two active regions. The main
objective of this work was to advance in the analysis of explosive
events by looking at the general properties of the samples rather than
studying individual cases.

The main conclusions from our work are that 
\begin{itemize}
\item even though we have significantly enlarged the sample size of
  explosive events by automating its detection, no clear and unique
  picture emerges, capable of describing the variety of line
  profiles encountered and described in Sect. \ref{propee};
\item current numerical simulations fail to explain the existance of
  explosive events line profiles with blue components much brighter
  than their red counterparts, and/or with maximum Doppler velocities in
  the red wing much larger than in the blue one;
\item the characteristics of explosive events near active regions as
  compared to their quiet Sun counterparts, and in particular their
  maximum velocities, cannot be explained only as the result of
  enhanced magnetic flux available for reconnection.
\end{itemize}

This work leaves many unanswered questions. In particular, we have not
analysed the time evolution of the explosive events, nor have we
correlated the position of explosive events near the two active
regions in the C~{\sc iv} and Ne~{\sc viii} lines. We intend to
explore these issues in the future, together with the latitude
dependence of the sample properties or the impact of the time
resolution by comparing this dataset with others available in the SOHO
data archive.

\begin{acknowledgements}

This research has made use of the Spanish Virtual Observatory
supported from the Spanish MEC through grant AyA2008-02156. We thank
the anonymous referee for greatly improving the readability of the
original manuscript.

\end{acknowledgements}

\bibliographystyle{aa}
\bibliography{sarro}

\begin{thebibliography}{28}
\expandafter\ifx\csname natexlab\endcsname\relax\def\natexlab#1{#1}\fi
\expandafter\ifx\csname url\endcsname\relax
  \def\url#1{{\tt #1}}\fi
\expandafter\ifx\csname urlprefix\endcsname\relax\def\urlprefix{URL }\fi

\bibitem[{{Bewsher} et~al.(2005){Bewsher}, {Innes}, {Parnell}, \&
  {Brown}}]{2005A&A...432..307B}
{Bewsher} D., {Innes} D.E., {Parnell} C.E., {Brown} D.S., Mar. 2005, \aap, 432,
  307

\bibitem[{{Brueckner} \& {Bartoe}(1983)}]{1983ApJ...272..329B}
{Brueckner} G.E., {Bartoe} J., Sep. 1983, \apj, 272, 329

\bibitem[{Castellanos \& Makarov(2006)}]{citeulike:781820}
Castellanos N.P., Makarov V.A., July 2006, J Neurosci Methods,
  \urlprefix\url{http://dx.doi.org/10.1016/j.jneumeth.2006.05.033}

\bibitem[{{Chae} et~al.(1998{\natexlab{a}}){Chae}, {Wang}, {Lee}, {Goode}, \&
  {Schuehle}}]{1998ApJ...497L.109C}
{Chae} J., {Wang} H., {Lee} C., {Goode} P.R., {Schuehle} U., Apr.
  1998{\natexlab{a}}, \apjl, 497, L109+

\bibitem[{{Chae} et~al.(1998{\natexlab{b}}){Chae}, {Wang}, {Lee}, {Goode}, \&
  {Schuhle}}]{1998ApJ...504L.123C}
{Chae} J., {Wang} H., {Lee} C., {Goode} P.R., {Schuhle} U., Sep.
  1998{\natexlab{b}}, \apjl, 504, L123+

\bibitem[{{Chae} et~al.(2000){Chae}, {Wang}, {Goode}, {Fludra}, \&
  {Sch{\"u}hle}}]{2000ApJ...528L.119C}
{Chae} J., {Wang} H., {Goode} P.R., {Fludra} A., {Sch{\"u}hle} U., Jan. 2000,
  \apjl, 528, L119

\bibitem[{{Chen} \& {Priest}(2006)}]{2006SoPh..238..313C}
{Chen} P.F., {Priest} E.R., Nov. 2006, \solphys, 238, 313

\bibitem[{{Dammasch} et~al.(1999){Dammasch}, {Wilhelm}, {Curdt}, \&
  {Hassler}}]{1999A&A...346..285D}
{Dammasch} I.E., {Wilhelm} K., {Curdt} W., {Hassler} D.M., Jun. 1999, \aap,
  346, 285

\bibitem[{{Dere}(1994)}]{1994AdSpR..14...13D}
{Dere} K.P., Apr. 1994, Advances in Space Research, 14, 13

\bibitem[{{Dere} et~al.(1989){Dere}, {Bartoe}, \&
  {Brueckner}}]{1989SoPh..123...41D}
{Dere} K.P., {Bartoe} J., {Brueckner} G.E., Mar. 1989, \solphys, 123, 41

\bibitem[{{Ding} et~al.(2010){Ding}, {Madjarska}, {Doyle}, \&
  {Lu}}]{2010A&A...510A.111D}
{Ding} J.Y., {Madjarska} M.S., {Doyle} J.G., {Lu} Q.M., Feb. 2010, \aap, 510,
  A111+

\bibitem[{{Doyle} et~al.(2005){Doyle}, {Ishak}, {Ugarte-Urra}, {Bryans}, \&
  {Summers}}]{2005A&A...439.1183D}
{Doyle} J.G., {Ishak} B., {Ugarte-Urra} I., {Bryans} P., {Summers} H.P., Sep.
  2005, \aap, 439, 1183

\bibitem[{Doyle et~al.(2006)Doyle, Popescu, \& Taroyan}]{Doyle:2006a}
Doyle J.G., Popescu M.D., Taroyan Y., 2006, Astronomy and Astrophysics,
  327--331

\bibitem[{Filzmoser et~al.(2008)Filzmoser, Maronna, \&
  Werner}]{Filzmoser:2008a}
Filzmoser P., Maronna R., Werner M., 2008, Computational Statistics and Data
  Analysis, 52, 1694

\bibitem[{{Heggland} et~al.(2009){Heggland}, {De Pontieu}, \&
  {Hansteen}}]{2009ApJ...702....1H}
{Heggland} L., {De Pontieu} B., {Hansteen} V.H., Sep. 2009, \apj, 702, 1

\bibitem[{Hyv{\"a}rinen(1999)}]{Hyvarinen:1999a}
Hyv{\"a}rinen A., 1999, IEEE Trans. on Neural Networks, 10, 626

\bibitem[{{Innes} et~al.(1997){Innes}, {Inhester}, {Axford}, \&
  {Wilhelm}}]{1997Natur.386..811I}
{Innes} D.E., {Inhester} B., {Axford} W.I., {Wilhelm} K., Apr. 1997, \nat, 386,
  811

\bibitem[{{Litvinenko} \& {Chae}(2009)}]{2009A&A...495..953L}
{Litvinenko} Y.E., {Chae} J., Mar. 2009, \aap, 495, 953

\bibitem[{{Madjarska} \& {Doyle}(2003)}]{2003A&A...403..731M}
{Madjarska} M.S., {Doyle} J.G., May 2003, \aap, 403, 731

\bibitem[{{Madjarska} et~al.(2009){Madjarska}, {Doyle}, \& {de
  Pontieu}}]{2009ApJ...701..253M}
{Madjarska} M.S., {Doyle} J.G., {de Pontieu} B., Aug. 2009, \apj, 701, 253

\bibitem[{{Mendoza-Torres} et~al.(2005){Mendoza-Torres}, {Torres-Papaqui}, \&
  {Wilhelm}}]{2005A&A...431..339M}
{Mendoza-Torres} J.E., {Torres-Papaqui} J.P., {Wilhelm} K., Feb. 2005, \aap,
  431, 339

\bibitem[{Ning et~al.(2004)Ning, Innes, \& Solanki}]{Ning:2004a}
Ning Z., Innes D.E., Solanki S.K., Jun. 2004, Astronomy and Astrophysics, 419,
  1141

\bibitem[{P{\'e}rez et~al.(1999)P{\'e}rez, Doyle, Erd{\'e}lyi, \&
  Sarro}]{Perez:1999a}
P{\'e}rez M., Doyle J., Erd{\'e}lyi R., Sarro L., 1999, Astronomy and
  Astrophysics, 342, 279

\bibitem[{{Peter} \& {Brkovi{\'c}}(2003)}]{2003A&A...403..287P}
{Peter} H., {Brkovi{\'c}} A., May 2003, \aap, 403, 287

\bibitem[{{Sarro} \& {Berihuete}(2008)}]{Sarro:2008a}
{Sarro} L.M., {Berihuete} A., Dec. 2008, In: American Institute of Physics
  Conference Series, vol. 1082 of American Institute of Physics Conference
  Series, 302--306

\bibitem[{{Teriaca} et~al.(1999){Teriaca}, {Banerjee}, \&
  {Doyle}}]{1999A&A...349..636T}
{Teriaca} L., {Banerjee} D., {Doyle} J.G., Sep. 1999, \aap, 349, 636

\bibitem[{Teriaca et~al.(2004)Teriaca, Banerjee, Falchi, Doyle, \&
  Madjarska}]{Teriaca:2004a}
Teriaca L., Banerjee D., Falchi A., Doyle J.G., Madjarska M.S., Dec. 2004,
  Astronomy and Astrophysics, 427, 1065

\bibitem[{{Wilhelm} et~al.(1995){Wilhelm}, {Curdt}, {Marsch}
  et~al.}]{1995SoPh..162..189W}
{Wilhelm} K., {Curdt} W., {Marsch} E., et~al., Dec. 1995, \solphys, 162, 189

\end{thebibliography}

\newpage

\Online

\clearpage 


\end{document}